\documentclass[aps,prd,superscriptaddress,groupedaddress,nofootinbib]{revtex4}  
\usepackage{dcolumn}   
\usepackage{bm}        
\usepackage[flushleft]{caption}

\usepackage{amsmath,amssymb,graphicx} 
\usepackage{epsfig}
\usepackage{pstricks}
\usepackage{multirow}
\usepackage[bookmarks]{hyperref}
\usepackage{subfig}
\usepackage{graphicx}
\usepackage{axodraw4j}
\usepackage{lipsum}

\usepackage[a4paper, left=2.5cm, right=2.5cm, top=3.5cm, bottom=2.5cm]{geometry}

\newcommand{\MET}{\diagup\hskip -8ptE_T}

\def\slash#1{\rlap{\hbox{$\mskip 1 mu /$}}#1}

\begin{document}
\title{Towards a model-independent approach 
to the analysis of interference effects  in pair production of new heavy quarks}

\author{D. Barducci}
\affiliation{\it School of Physics and Astronomy, University of Southampton,\\ Highfield, Southampton SO17 1BJ, UK}
\affiliation{Particle Physics Department, Rutherford Appleton Laboratory,  
       \\Chilton, Didcot, Oxon OX11 0QX, UK} 
\author{A. Belyaev}
\affiliation{\it School of Physics and Astronomy, University of Southampton,\\ Highfield, Southampton SO17 1BJ, UK}
\affiliation{Particle Physics Department, Rutherford Appleton Laboratory,  
       \\Chilton, Didcot, Oxon OX11 0QX, UK} 
\author{J. Blamey}
\affiliation{\it School of Physics and Astronomy, University of Southampton,\\ Highfield, Southampton SO17 1BJ, UK}
\author{S. Moretti}
\affiliation{\it School of Physics and Astronomy, University of Southampton,\\ Highfield, Southampton SO17 1BJ, UK}
\affiliation{Particle Physics Department, Rutherford Appleton Laboratory,  
       \\Chilton, Didcot, Oxon OX11 0QX, UK} 
\author{L. Panizzi}
\affiliation{\it School of Physics and Astronomy, University of Southampton,\\ Highfield, Southampton SO17 1BJ, UK}
\affiliation{Particle Physics Department, Rutherford Appleton Laboratory,  
       \\Chilton, Didcot, Oxon OX11 0QX, UK} 
\author{H. Prager}
\affiliation{\it School of Physics and Astronomy, University of Southampton,\\ Highfield, Southampton SO17 1BJ, UK}
\affiliation{\'Ecole Normale Sup\'erieure de Lyon, Universit\'e de Lyon, 46 all\'ee d'Italie, 69364 Lyon cedex 07, France}

\date\today

\begin{abstract}

We propose a model independent approach for the analysis of interference effects 
in the process of QCD pair production of new heavy quarks of different species
that decay into Standard Model particles, 
including decays via flavour changing neutral currents.
By adopting as ansatz a simple analytical formula 
we show that one can accurately describe the interference 
between two different such particle pairs leading to the same final state using information about  masses, total widths and couplings.
A study of the effects on differential distributions is
also performed showing that, when interference plays a relevant role, the distributions of the full
process can be obtained by a simple rescaling of the distributions of either quark contributing to the
interference term.
We also present the range of validity of the analytical expression that we have found.

\end{abstract}

\maketitle

\section{Introduction}

The discovery of a Higgs boson \cite{Aad:2012tfa,Chatrchyan:2012ufa} has essentially excluded  a fourth generation of chiral quarks~\cite{Djouadi:2012ae,Eberhardt:2012gv}, thus shifting the focus of  
new heavy quark searches towards vector-like quarks ($Q_V$s). The latter are heavy spin 1/2 particles that 
transform as triplets under colour and whose left- and right-handed couplings have the same
Quantum Chromo-Dynamics (QCD)  and Electro-Weak (EW) quantum numbers.
These states are predicted by various theoretical models (composite Higgs models \cite{Dobrescu:1997nm,Chivukula:1998wd,He:2001fz,Hill:2002ap,Agashe:2004rs,Contino:2006qr,Barbieri:2007bh,Anastasiou:2009rv}, models with extra dimensions, little Higgs models \cite{ArkaniHamed:2002qy,Schmaltz:2005ky}, models with gauging of the flavour group \cite{Davidson:1987tr,Babu:1989rb,Grinstein:2010ve,Guadagnoli:2011id},  non-minimal
supersymmetric  extensions of the Standard Model (SM) \cite{Moroi:1991mg,Moroi:1992zk,Babu:2008ge,Martin:2009bg,Graham:2009gy,Martin:2010dc}, Grand Unified Theories  \cite{Rosner:1985hx,Robinett:1985dz}) and can be observed in a large number of final states, depending on how they interact with SM particles (see for example \cite{AguilarSaavedra:2009es,Okada:2012gy,DeSimone:2012fs,Buchkremer:2013bha,Aguilar-Saavedra:2013qpa} for general reviews).

Usually experimental searches for vector-like quarks adopt a phenomenological approach, assuming that only one new 
$Q_V$ state is present beyond the SM and, in order to be as model independent as possible, searches usually consider QCD pair production, although very recently
 single production has also been explored \cite{Aad:2013rna}. Most models, however, predict in general the existence of a new \textit{quark sector}, which implies the presence of more than one new coloured state, some of which
being possibly degenerate or nearly degenerate. If two or more quarks
 of a given model can decay to the same final 
state, interference effects should be considered in order to correctly evaluate the total cross section
 and the kinematical distributions of the signal. Current bounds on the masses of new states obtained assuming the presence of only one new particle cannot be easily reinterpreted in more complex scenarios containing more than one new quark, unless interference
 effects in the total cross section and kinematical distributions are taken into account.

We show that this can be done through a simple formula, which enables one to correctly model such interference effects
at both inclusive and exclusive levels. The plan of the paper is as follows. In the next section we describe the procedure while in the following one we  present our numerical results. 
Then, we conclude.

\section{Analytical  Estimation of the Interference effects for pair vector-like quarks production}

\subsection{Analytical ``master formula" for the interference}

We will assume throughout the analysis that the new heavy quarks undergo two-body decays to SM particles and we will not consider chain decays of heavy quarks into other new states, possibly including dark matter candidates. This
approach is generally valid for  models in which the new quarks interact with the SM ones only through Yukawa couplings.
Therefore, the new heavy quarks can decay into either SM gauge
 bosons or the Higgs boson and ordinary quarks. We will assume that flavour changing neutral currents are present and therefore decays such as $t^\prime\to Zt$ and $t^\prime\to Ht$ are allowed, alongside $t^\prime\to W^+ b$. This is consistent with the embedding of new $Q_V$s in extensions of the SM. 
If more than one $Q_V$ species  is present in the model, then there are two ways to obtain a given final state:
\begin{itemize}
 \item[A.]  $Q_V^i$ quarks have the  same charge, so a $Q_V^i \bar{Q}_V^i$ pair  decays into the same final state, e.g., \\
 $t^\prime_{1,2}\bar{t}^\prime_{1,2} \to  W^+W^-b \bar b (W^+Zb\bar t)$;
 \item[B.]  $Q_V^i$ quarks have  different charges  but after decay  their  pair leads to the same final state, e.g.,\\
 $b^\prime \bar b^\prime \to (t W^-)(\bar t W^+)$ and $X_{5/3} \bar X_{5/3} \to (t W^+)(\bar t W^-)$.
\end{itemize}

We have verified that, while the interference in case B can be safely neglected when the masses of the vector-like quarks are much larger than the masses of the decay products (which is usually the case), because of the largely different kinematics of the final states, case A has to be considered carefully. 
It is worth mentioning that, for the classes of models under consideration,
we  have   quarks of identical charge and with couplings to the same particles, so that 
the effects of  the mixing between such quarks at loop level
could be important and should (eventually) be taken into account. These effects are model-dependent though 
and involve computation of loops that may contain states belonging to new sectors (e.g., new gauge bosons). 
In this paper we  assume  that these effects can be computed 
and that particle wave-function as well as Feynman rules are already formulated 
for mass-eigenstates, i.e., the masses and widths that we will be using are those obtained after computing the rotations of the states due to the one-loop mixing terms, so that interference effects can then be explored in a model-independent way.

The measure  of the  interference between $Q_V^i$ and  $Q_V^j$ pairs 
of species $i$ and $j$ decaying into the same final state can be defined by the following simple expression
\begin{equation}
F_{ij}= 
\frac{\sigma^{\rm int}_{ij}}{\sigma_i+\sigma_j}
=
\frac{\sigma^{\rm tot}_{ij}-(\sigma_i+\sigma_j)}{\sigma_i+\sigma_j}
=
\frac{\sigma^{\rm tot}_{ij}}{\sigma_i+\sigma_j}-1
\label{orderparameter}
\end{equation}
where
$\sigma^{\rm tot}_{ij}$ is the total cross section
of  $Q_V^i$ and  $Q_V^j$  pair production including their interference, the
$\sigma_{i,j}$s are their individual production rates
while $\sigma^{\rm int}_{ij}$ represents the value of the interference.

The interference  term $F_{ij}$ ranges from $-1$ to 1. Completely constructive interference is obviously achieved when $\sigma^{\rm int}_{ij}=\sigma_i+\sigma_j$, while completely destructive interference is obtained when $\sigma^{\rm int}_{ij}=-(\sigma_i+\sigma_j)$.

It is known that, under very general hypotheses, the couplings of $Q_V$s with SM quarks are dominantly chiral and that the chirality of the coupling depends on the $Q_V$ representation under $SU(2)$~\cite{delAguila:1982fs,delAguila:2000rc,AguilarSaavedra:2009es,DeSimone:2012fs,Buchkremer:2013bha}. If the $Q_V$ belongs to a half-integer representation (doublets, quadruplets, \dots) couplings are dominantly right-handed while, if the $Q_V$ belongs
 to an integer representation (singlets, triplets, \dots) couplings are mostly left-handed. This feature is valid for a wide range of hypotheses about the mixing between $Q_V$s and SM quarks and between $Q_V$s themselves. However, if Yukawa couplings
 between $Q_V$s and the Higgs boson are large, it is possible to achieve couplings with non-dominant chiralities.

Our results about the analysis of interference effects can be applied in both cases, therefore, we divide our 
study in two parts. Firstly, we show the results for the interference of two $t^\prime$s with the same chiral couplings. Then we generalise the analysis to the case where the couplings of the heavy quarks do not exhibit a dominant chirality.

We would  now like to make the ansatz that, in case of chiral new quarks $i$ and assuming small  $\Gamma_i/m_i$ values, the interference is proportional to the couplings of the new quarks to the final state particles and to the integral of the {\it scalar} part of the propagator. The range of validity of the ansatz in terms of the $\Gamma_i/m_i$ ratio is explored in a subsequent section.

If the couplings are chiral for both heavy quarks and the chirality is the same we have
\begin{equation}
 \label{eq:intpart}
 \sigma^{\rm int}_{ij} \propto 2 Re\left[g_{i1} g_{j1}^* g_{i2}^* g_{j2} \left(\int_{-\infty}^{+\infty} d q^2 \mathcal{P}_i\mathcal{P}_j^*\right)^2\right]
\end{equation}
where 1 and 2 refer to the two decay branches (1 corresponding to the quark branch and 2 to the antiquark branch)
while the scalar part of the propagator for any new quark $i$ is given by
\begin{equation}
\mathcal{P}_i=\frac{1}{q^2-m_i^2+im_i\Gamma_i}.
\end{equation}
The cross section for pair production of  species $i$ only is 
\begin{equation}
  \label{eq:prodpart}
 \sigma_{i} \propto |g_{i1}|^2 |g_{i2}|^2 \left(\int d q^2 \mathcal{P}_i\mathcal{P}_i^*\right)^2
\end{equation}
and an analogous expression can be written for species $j$.

Therefore, the analytical  expression which should describe the interference in the case of chiral $Q_V$
pair production of species $i$ and $j$ followed by their decay into the same final state, is given by
\begin{equation}
\label{eq:int1}
\kappa_{ij}=\frac{2 Re\left[g_{i1} g_{j1}^* g_{i2}^* g_{j2} \left(\int \mathcal{P}_i\mathcal{P}_j^*\right)^2\right]}{|g_{i1}|^2 |g_{i2}|^2 \left(\int \mathcal{P}_i\mathcal{P}_i^*\right)^2+|g_{j1}|^2 |g_{j2}|^2 \left(\int \mathcal{P}_j\mathcal{P}_j^*\right)^2}.
\end{equation}
Ultimately, $\kappa_{ij}$ should closely describe the true value of the interference term $F_{ij}$ from  Eq.~(\ref{orderparameter})
if the ansatz is correct.

After integration $\kappa_{ij}$ takes the following form:
\begin{equation}
\label{eq:int2}
\kappa_{ij}=\frac{8 Re[g_{i1} g_{j1}^* g_{i2}^* g_{j2}] m_i^2 m_j^2 \Gamma_i^2 \Gamma_j^2}{|g_{j1}|^2|g_{j2}|^2m_i^2\Gamma_i^2+|g_{i1}|^2|g_{i2}|^2m_j^2\Gamma_j^2} \frac{(m_i\Gamma_i+m_j\Gamma_j)^2-(m_i^2 - m_j^2)^2}{\left((m_i\Gamma_i+m_j\Gamma_j)^2+(m_i^2 - m_j^2)^2\right)^2}.
\end{equation}

The previous expression can be generalised when the chirality of the coupling is not predominantly left or right. In the approximation in which the final states are massless (in practice, neglecting the top mass) only four sub-diagrams give a 
non-zero contribution, the ones corresponding to considering the following combinations of chiralities: $q_1^\prime,q_2^\prime,\bar q_1^\prime,\bar q_2^\prime$=$L,L,L,L$ or $L,L,R,R$ or $ R,R,L,L$ or $R,R,R,R$. If the masses of the final state objects cannot be neglected, the non-zero combinations would be 16 because any combination of $q^\prime_1$ would interfere with any combination of $q^\prime_2$, though interferences involving $LR$ or $RL$ flipping are suppressed by the mass of the quarks in the final state. Analogously to the previous case, we have numerically proven that neglecting the masses of the final states is a reasonable assumption in the range of $Q_V$ masses still allowed by experimental data, hence we will consider the final state quarks as massless.

The expression in Eq. (\ref{eq:int1}) can therefore be rewritten in the following way:
\begin{equation}
\kappa^{ab}_{ij}=\frac{2 Re\left[g^a_{i1} g_{j1}^{a*} g^{b*}_{i2} g_{j2}^{b} \left(\int \mathcal{P}_i\mathcal{P}_j^*\right)^2\right]}{|g^a_{i1}|^2 |g^b_{i2}|^2 \left(\int \mathcal{P}_i\mathcal{P}_i^*\right)^2+|g^a_{j1}|^2 |g^b_{j2}|^2 \left(\int \mathcal{P}_j\mathcal{P}_j^*\right)^2}=\frac{\mathcal{N}_{ij}^{ab}}{\mathcal{D}_{ij}^{ab}}, \qquad ab=LL,LR,RL,RR.
\end{equation}
After summing over all allowed topologies, we obtain the generalisation of Eq.(\ref{eq:int2}) as:
\begin{equation}
\label{eq:gen1}
\kappa^{gen}_{ij}=
\frac{\sum_{a,b=L,R}2 Re\left[g^a_{i1} g_{j1}^{a*} g^{b*}_{i2} g_{j2}^{b} \left(\int \mathcal{P}_i\mathcal{P}_j^*\right)^2\right]}{\sum_{a,b=L,R}|g^a_{i1}|^2 |g^b_{i2}|^2 \left(\int \mathcal{P}_i\mathcal{P}_i^*\right)^2+|g^a_{j1}|^2 |g^b_{j2}|^2 \left(\int \mathcal{P}_j\mathcal{P}_j^*\right)^2}=\frac{\sum_{ab}\kappa_{ij}^{ab}\mathcal{D}_{ij}^{ab}}{\sum_{ab}\mathcal{D}_{ij}^{ab}},
\end{equation}
which,  after integration, becomes
\begin{eqnarray}
\label{eq:gen2}
\kappa^{gen}_{ij}&=&
\frac{8 Re[(g_{i1}^L g_{j1}^{L*} + g_{i1}^R g_{j1}^{R*})(g_{i2}^{L*}g_{j2}^L + g_{i2}^{R*}g_{j2}^R)] m_i^2 m_j^2 \Gamma_i^2 \Gamma_j^2}{\left((|g_{j1}^L|^2+|g_{j1}^R|^2)(|g_{j2}^L|^2+|g_{j2}^R|^2\right)m_i^2\Gamma_i^2+ \left( (|g_{i1}^L|^2+|g_{i1}^R|^2) (|g_{i2}^L|^2+|g_{i2}^R|^2) \right) m_j^2\Gamma_j^2}\cdot\nonumber \\ &&\frac{(m_i\Gamma_i+m_j\Gamma_j)^2-(m_i^2 - m_j^2)^2}{\left((m_i\Gamma_i+m_j\Gamma_j)^2+(m_i^2 - m_j^2)^2\right)^2}.
\end{eqnarray}


\subsection{Region of validity of the approximation}

When considering the production and decay of different heavy quarks which couple to the same SM  particles, interference at tree level is not the only one which should potentially be taken into account. Quarks with same quantum numbers can mix at loop level too, 
which results into the respective mixing matrix of the one-loop corrected propagators
and their corresponding interference.
Mass and width eigenstates can be obtained by diagonalising the respective matrices, but the rotations are in general different for these two matrices, therefore mass and width eigenstates may be misaligned. A careful treatment of all such mixing effects is beyond the scope of this analyis but, in order to be able to apply our results, it is crucial to understand when the mixing effect can be neglected.

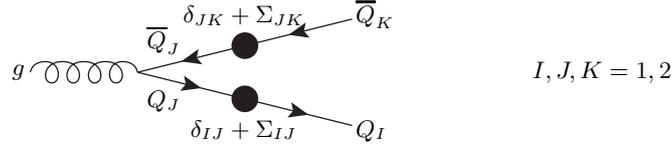
\begin{figure}[htb]
\begin{center}
 \begin{picture}(250,44)(0,0)
    \SetColor{Black}
    \Gluon(10,22)(50,22){4}{4.5}
    \Text(8,22)[rc]{$g$}
    \Line[arrow,arrowpos=0.5,arrowwidth=2](90,32)(50,22)
    \Text(60,29)[cb]{$\overline{Q}_J$}
    \Line[arrow,arrowpos=0.5,arrowwidth=2](50,22)(90,12)
    \Text(60,15)[ct]{$Q_J$}
    \Vertex(90,32){5}
    \Text(90,40)[cb]{$\delta_{JK}+\Sigma_{JK}$}
    \Vertex(90,12){5}
    \Text(90,4)[ct]{$\delta_{IJ}+\Sigma_{IJ}$}
    \Line[arrow,arrowpos=0.5,arrowwidth=2](130,42)(90,32)
    \Text(132,44)[lc]{$\overline{Q}_K$}
    \Line[arrow,arrowpos=0.5,arrowwidth=2](90,12)(130,2)
    \Text(132,0)[lc]{$Q_I$}
    \Text(250,22)[rc]{$I,J,K=1,2$}
  \end{picture}
\end{center}
\caption{Pair production of two heavy quarks $Q_1$ and $Q_2$, including loop mixing.}
\label{fig:pairprodwithloops}
\end{figure}

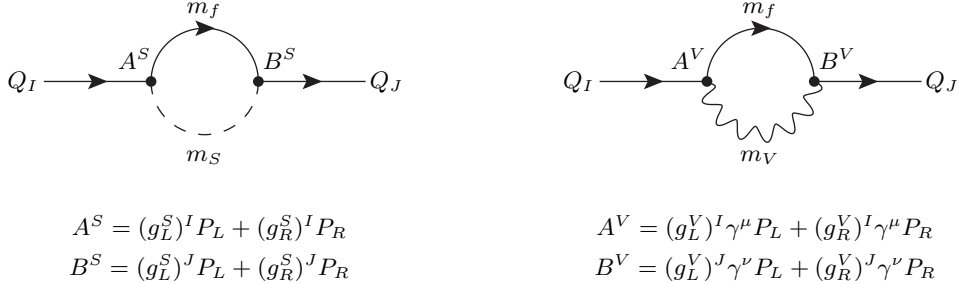
\begin{figure}[htb]
\begin{minipage}{.45\textwidth}
\begin{center}
\begin{center}
 \begin{picture}(150,120)(0,-10)
    \SetColor{Black}
    \Line[arrow,arrowpos=0.5,arrowwidth=2](10,70)(50,70)
    \Text(8,70)[rc]{$Q_I$}
    \Vertex(50,70){2}
    \Text(50,75)[rb]{$A^S$}
    \Arc[arrow,arrowpos=0.5,arrowlength=5,arrowwidth=2,arrowinset=0.2,clock](70,70)(20,-180,-360)
    \Text(70,94)[cb]{$m_f$}
    \Vertex(90,70){2}
    \Text(92,75)[lb]{$B^S$}
    \Arc[dash,dashsize=5,clock](70,70)(20,-0,-180)
    \Text(70,44)[ct]{$m_S$}
    \Line[arrow,arrowpos=0.5,arrowwidth=2](90,70)(130,70)
    \Text(132,70)[lc]{$Q_J$}
    \Text(72,15)[cc]{$A^{S}=(g_L^S)^I P_L + (g_R^S)^I P_R$}
    \Text(72,0)[cc]{$B^{S}=(g_L^S)^J P_L + (g_R^S)^J P_R$}
  \end{picture}
\end{center}
\end{center}
\end{minipage}
\begin{minipage}{.45\textwidth}
\begin{center}
 \begin{picture}(150,120)(0,-10)
    \SetColor{Black}
    \Line[arrow,arrowpos=0.5,arrowwidth=2](10,70)(50,70)
    \Text(8,70)[rc]{$Q_I$}
    \Vertex(50,70){2}
    \Text(50,75)[rb]{$A^V$}
    \Arc[arrow,arrowpos=0.5,arrowlength=5,arrowwidth=2,arrowinset=0.2,clock](70,70)(20,-180,-360)
    \Text(70,94)[cb]{$m_f$}
    \Vertex(90,70){2}
    \Text(92,75)[lb]{$B^V$}
    \PhotonArc[clock](70,70)(20,-0,-180){3}{7}
    \Text(70,44)[ct]{$m_V$}
    \Line[arrow,arrowpos=0.5,arrowwidth=2](90,70)(130,70)
    \Text(132,70)[lc]{$Q_J$}
    \Text(72,15)[cc]{$A^{V}=(g_L^V)^I \gamma^\mu P_L + (g_R^V)^I \gamma^\mu P_R$}
    \Text(72,0)[cc]{$B^{V}=(g_L^V)^J \gamma^\nu P_L + (g_R^V)^J \gamma^\nu P_R$}
  \end{picture}
\end{center}
\end{minipage}
\caption{Loop topologies for corrections to quark propagators. The particles  in the loop can be any fermion, vector or scalar which are present in the model under consideration.}
\label{fig:loopnotation}
\end{figure}

Let us consider the structure of the interference terms for the process of QCD pair production of two heavy quarks, $Q_1$ and $Q_2$, including the one-loop corrections to the quark propagators. From now on we will consider only the imaginary part of the quark self-energies, that give the corrections to the quark widths, and we will assume real couplings for simplicity. A more detailed treatment of mixing effects under general assumptions in heavy quark pair production will be performed in a dedicated analysis~\cite{mixingVLQ}. Considering only the case of s-channel exchange of the gluon for simplicity, and still not including the decays of the heavy quarks, the amplitude of the process depicted in Fig.\ref{fig:pairprodwithloops} is:
\begin{equation}
 \mathcal{M} = \bar{u}_I (\delta_{IJ}+\Sigma_{IJ}) P^+_J V^\sigma P^-_J (\delta_{JK}+\Sigma_{JK}) v_K \mathcal{M}^P_\sigma \quad\mbox{with}\quad I,J,K=1,2
\end{equation}
where the QCD amplitude terms and colour structure have been factorised into the vertex $V^\sigma$ and the term $\mathcal{M}^P_\sigma$, the propagators of the quark and antiquarks are $P^+$ and $P^-$, respectively, and $\Sigma$ represents the loop insertions. The loop contributions depend on the particle content of the model and therefore cannot be evaluated in a model independent way. However, it is straightforward to determine the structure of the loops by noticing that the only allowed topologies are fermion-scalar (fS) and fermion-vector (fV), see Fig.\ref{fig:loopnotation}.

These topologies can be evaluated for general masses and couplings of the particles  in the loops, and therefore the most general structure of the loop insertion is:
\begin{equation}
 \Sigma_{IJ} = \sum_{\mbox{fS loops}} \Sigma_{IJ}^{fS} + \sum_{\mbox{fV loops}} \Sigma_{IJ}^{fV}
\end{equation}
where, in Feynman gauge and adopting the Passarino-Veltman functions $B_0$ and $B_1$:
\begin{eqnarray}
 \Sigma_{IJ}^{fS} &=& \left((g^S_L)^I (g^S_L)^J m_f B_0(p^2,m_f^2,m_S^2) + (g^S_R)^I (g^S_L)^J \slash p B_1(p^2,m_f^2,m_S^2) \right) P_L + L \leftrightarrow R, \\
 \Sigma_{IJ}^{fV} &=& \left(4(g^V_R)^I (g^V_L)^J m_f B_0(p^2,m_f^2,m_V^2) - 2 (g^V_L)^I (g^V_L)^J \slash p B_1(p^2,m_f^2,m_V^2) \right) P_L + L \leftrightarrow R.
\end{eqnarray}
When $I=J$, the loop contributions correspond to a correction to the diagonal quark propagators while, when $I\neq J$, the loops correspond to the off-diagonal mixing between the quarks.
Without loosing generality, let us consider  the $I,K=1,2$ case, for which
we can  define two amplitude matrices, corresponding to production of the quarks $J=1$ and $J=2$ that, through the loop-corrected propagators, become quarks $I,K=1,2$. 

The amplitude matrices are:
\begin{eqnarray}
\mathcal{M}_{J=1} &=& \left(
\begin{array}{cc}
 \bar{u}_1 (1+\Sigma_{11}) P^+_1 V^\sigma P^-_1 (1+\Sigma_{11}) v_1 \mathcal{M}^P_\sigma & \bar{u}_1 (1+\Sigma_{11}) P^+_1 V^\sigma P^-_1 \Sigma_{12} v_2 \mathcal{M}^P_\sigma \\
 \bar{u}_2 \Sigma_{21} P^+_1 V^\sigma P^-_1 (1+\Sigma_{11}) v_1 \mathcal{M}^P_\sigma & \bar{u}_2 \Sigma_{21} P^+_1 V^\sigma P^-_1 \Sigma_{12} v_2 \mathcal{M}^P_\sigma
\end{array}
\right), \\
\mathcal{M}_{J=2} &=& \left(
\begin{array}{cc}
 \bar{u}_1 \Sigma_{12} P^+_2 V^\sigma P^-_2 \Sigma_{21} v_1 \mathcal{M}^P_\sigma & \bar{u}_1 \Sigma_{12} P^+_2 V^\sigma P^-_2 (1+\Sigma_{22}) v_2 \mathcal{M}^P_\sigma \\
 \bar{u}_2 (1+\Sigma_{22}) P^+_2 V^\sigma P^-_2 \Sigma_{21} v_1 \mathcal{M}^P_\sigma & \bar{u}_2 (1+\Sigma_{22}) P^+_2 V^\sigma P^-_2 (1+\Sigma_{22}) v_2 \mathcal{M}^P_\sigma
\end{array}
\right). 
\end{eqnarray}
The interference contribution of the cross-section can be obtained by contracting elements of one matrix with elements of the other matrix. Some interesting consequences can be derived from the structure of these matrices.
\begin{enumerate}
 \item It is possible to construct four interference terms by contracting elements with same indices (e.g. $\mathcal{M}_{J=1}|_{(1,1)}$ with $\mathcal{M}_{J=2}|_{(1,1)}$) due to the fact that the quarks in the final state are the same. At lowest order these interference terms will always contain two off-diagonal loop corrections.
 \item Any element of one matrix can be contracted with any element of the other matrix only when considering also the decays of the quarks, there fixing specific decay channels for the quark and antiquark branches. This way it is possible to obtain 16 interference combinations. The order of the interference term and the number of off-diagonal mixing contributions, however, will not always be the same, depending on the contraction. In particular, when contracting the element (1,1) of the $\mathcal{M}_{J=1}$ matrix with the element (2,2) of the $\mathcal{M}_{J=2}$ matrix, there are no off-diagonal loop mixings involved and the contraction after the quark decays will be given by a pure tree level contribution plus diagonal loop corrections while, when contracting the element (2,2) of the $\mathcal{M}_{J=1}$ matrix with the element (1,1) of the $\mathcal{M}_{J=2}$ matrix, there are 4 off-diagonal loop mixings involved, 
 so that this process, which has mixing terms to a higher power,  is expected to be suppressed.
\end{enumerate}
It is interesting to notice that, in the case of same-element contractions before quark decays (case 1), the order of the process is the same as in the case of contractions after quark decays of the element (1,1) of the $\mathcal{M}_{J=1}$ matrix with the element (2,2) of the $\mathcal{M}_{J=2}$ matrix (case 2). Therefore, the 4 interference contributions of case 1 can be competitive with the tree-level interference term after quark decay. However, if the off-diagonal contributions to the mixing matrix are negligible with respect to the diagonal elements, the two amplitude matrices reduce to:
\begin{eqnarray}
\mathcal{M}_{J=1} &\simeq& \left(
\begin{array}{cc}
 \bar{u}_1 (1+\Sigma_{11}) P^+_1 V^\sigma P^-_1 (1+\Sigma_{11}) v_1 \mathcal{M}^P_\sigma & 0 \\
 0 & 0
\end{array}
\right), \\ 
\mathcal{M}_{J=2} &\simeq& \left(
\begin{array}{cc}
 0 & 0 \\
 0 & \bar{u}_2 (1+\Sigma_{22}) P^+_2 V^\sigma P^-_2 (1+\Sigma_{22}) v_2 \mathcal{M}^P_\sigma
\end{array}
\right).
\end{eqnarray}
In this case the  same-element contraction of case 1 do not enter the determination of the interference terms and the lowest order contribution is given by contracting the only non-zero elements of the matrices at tree level after the decays of the quarks. In other words, 
the analytical description of the interference
developed in the previous section can only be applied in the case of suppressed or negligible 
mixing between the heavy quarks. One should note that the requirement of suppression of 
off-diagonal mixing can be potentially quite restrictive, since it will take place
in case of cancellation of loop contributions in the kinematic $p^2\simeq M_Q^2$ region where the couplings of the heavy quarks are chosen to compensate the different values of the loop integrals. The verification of such a case is eventually model-dependent and requires computing the mixing matrix structure, which in turn depends on the particle content of the model. 
For example in case of the off-diagonal contributions to the propagators of two top partners $T_1$ and $T_2$ that only couple to the third family of SM quarks and with all SM gauge bosons and the Higgs boson,
and requiring their sum to be suppressed with respect to the sum of the diagonal contributions,
we obtain the following relation:
\begin{equation}
  \Sigma_{IJ}=\Sigma_{IJ}^{tH}+\Sigma_{IJ}^{tZ}+\Sigma_{IJ}^{bW}+\Sigma_{IJ}^{tG^0}+\Sigma_{IJ}^{bG^+}\ll\{\Sigma_{II},\Sigma_{JJ}\}
\end{equation}
with $I,J=1,2$ and $I\neq J$. The suppression of the off-diagonal contribution depends on all the masses and couplings involved, plus it also depends on the $p^2$ of the external heavy quarks. However, if it is possible to find coupling configurations which satisfy the relation for a large $p^2$ region, our approach can be safely adopted. A detailed numerical treatment of this relation for different particle contents and coupling values is beyond the scope of this preliminary analysis, but it will be developed in a future one \cite{mixingVLQ}. 
It is also interesting to notice that, if the mass and width eigenvalues are not misaligned, it is possible to diagonalise the matrix of the propagators and define new states with definite mass and eigenstates. In this case it is possible to consider the exact amplitude matrix,
\begin{eqnarray}
\mathcal{M}_{J={1^\prime}} &=& \left(
\begin{array}{cc}
 \bar{u}_{1^\prime} P^+_{1^\prime} V^\sigma P^-_{1^\prime} v_{1^\prime} \mathcal{M}^P_\sigma & 0 \\
 0 & 0
\end{array}
\right), \\ 
\mathcal{M}_{J={2^\prime}} &=& \left(
\begin{array}{cc}
 0 & 0 \\
 0 & \bar{u}_{2^\prime} P^+_{2^\prime} V^\sigma P^-_{2^\prime} v_{2^\prime} \mathcal{M}^P_\sigma
\end{array}
\right),
\end{eqnarray}
then compute the tree-level interference after the decays of the quarks with the method developed in the previous section, but considering quarks with loop-corrected masses and widths. Again, this is a specific situation, but it is a further case when the relations stuided in this paper can be applied.

\section{Numerical results}

\subsection{Total cross section}

\begin{figure}[htb]
\begin{center}
 \begin{picture}(160,100)(0,0)
    \SetWidth{2.0}
    \SetColor{Black}
    \Line[arrow,arrowpos=0.5,arrowwidth=2](10,60)(40,50)
    \Text(8,60)[rc]{\small{$p$}}
    \Line[arrow,arrowpos=0.5,arrowwidth=2](10,20)(40,30)
    \Text(8,20)[rc]{\small{$p$}}
    \SetWidth{1.0}
    \Line[arrow,arrowpos=0.5,arrowwidth=2](60,50)(90,60)
    \Text(75,60)[cb]{\small{$t^\prime$}}
    \Line[arrow,arrowpos=0.5,arrowwidth=2](60,30)(90,20)
    \Text(75,20)[ct]{\small{$\bar t^\prime$}}
    \Line[arrow,arrowpos=0.5,arrowwidth=2](90,60)(120,70)
    \Text(122,70)[lc]{\small{$b$}}
    \Line[arrow,arrowpos=0.5,arrowwidth=2](90,20)(120,10)
    \Text(122,10)[lc]{\small{$\bar t$}}
    \Photon(120,50)(90,60){2}{4.5}
    \Text(122,50)[lc]{\small{$W^+$}}
    \Photon(90,20)(120,30){2}{4.5}
    \Text(122,30)[lb]{\small{$Z$}}
    \GOval(50,40)(15,15)(0){0.882}
  \end{picture}
\end{center}
\caption{Pair production of a pair of $t^\prime$ $Q_V$s and subsequent decay into a $bW^+\bar t Z$ final state. }
\label{fig:prod-decay}
\end{figure}
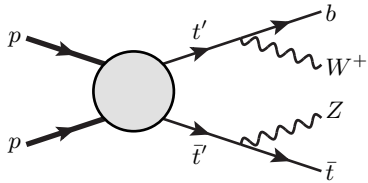

We first consider the production and decay rates of two $t^\prime$s pairs decaying into $W^+b$ and $Z \bar t$, see Fig.~\ref{fig:prod-decay}, i.e., we consider the $2\to4$ process
\begin{equation}
\label{eq:proc}
p p \rightarrow t_i^\prime \bar t_i^\prime \rightarrow W^+ b Z \bar t, \qquad i=1,2,
\end{equation}
with the chirality of the couplings being the same for the two states.
This process has been chosen to provide a concrete example; in general, vector-like quarks can also decay into the Higgs boson, but we have fixed a specific final state to perform the simulations. Selecting different final states involving decays into Higgs would give analogous results.

\begin{figure}[htb]
\begin{minipage}{.48\textwidth}
\hskip 20pt \bf Chiral couplings
\end{minipage}\hfill
\begin{minipage}{.48\textwidth}
\hskip 20pt \bf General couplings
\end{minipage}
\vskip 10pt
\epsfig{file=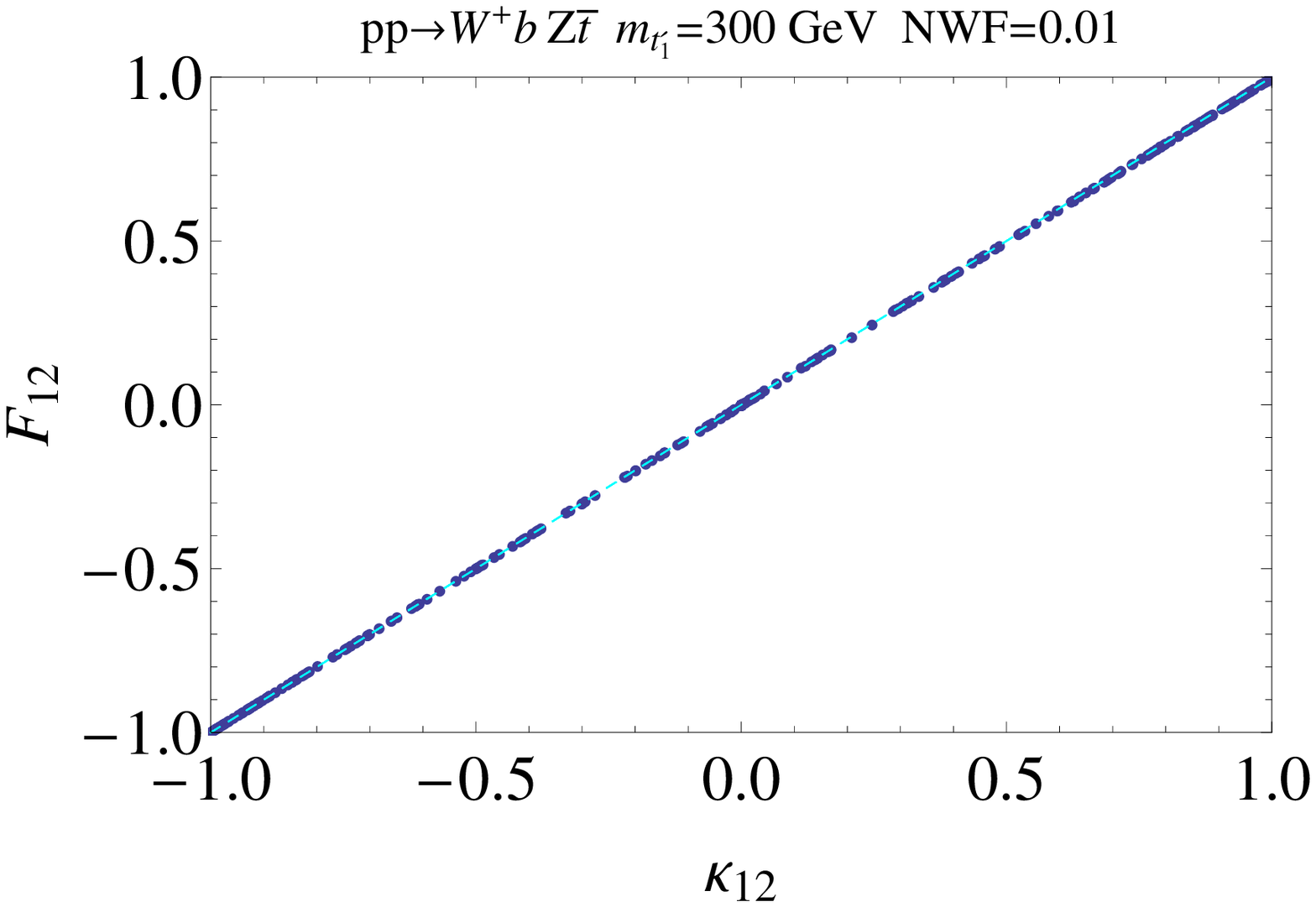, width=0.5\textwidth}%
\epsfig{file=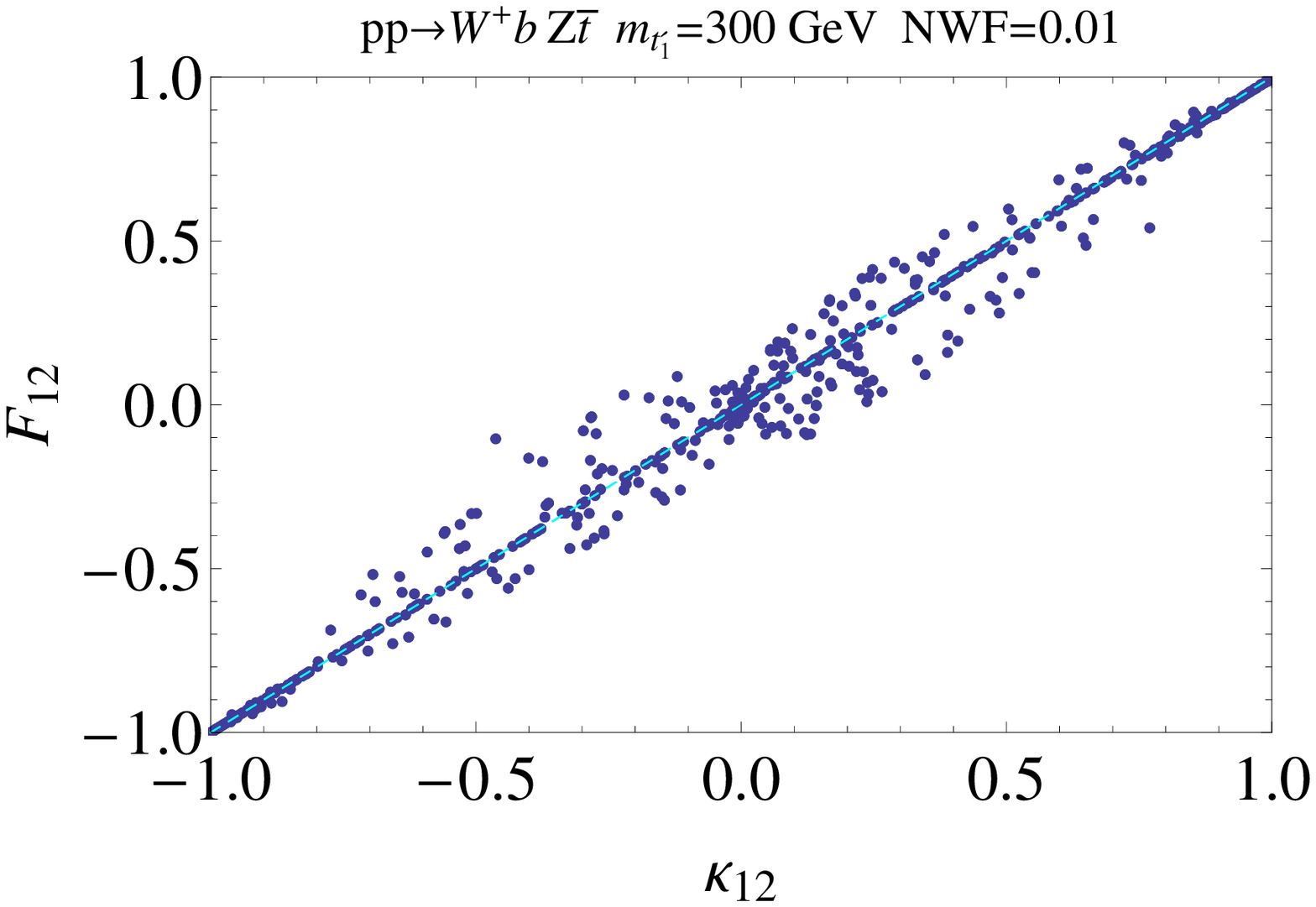, width=0.5\textwidth}\\
\epsfig{file=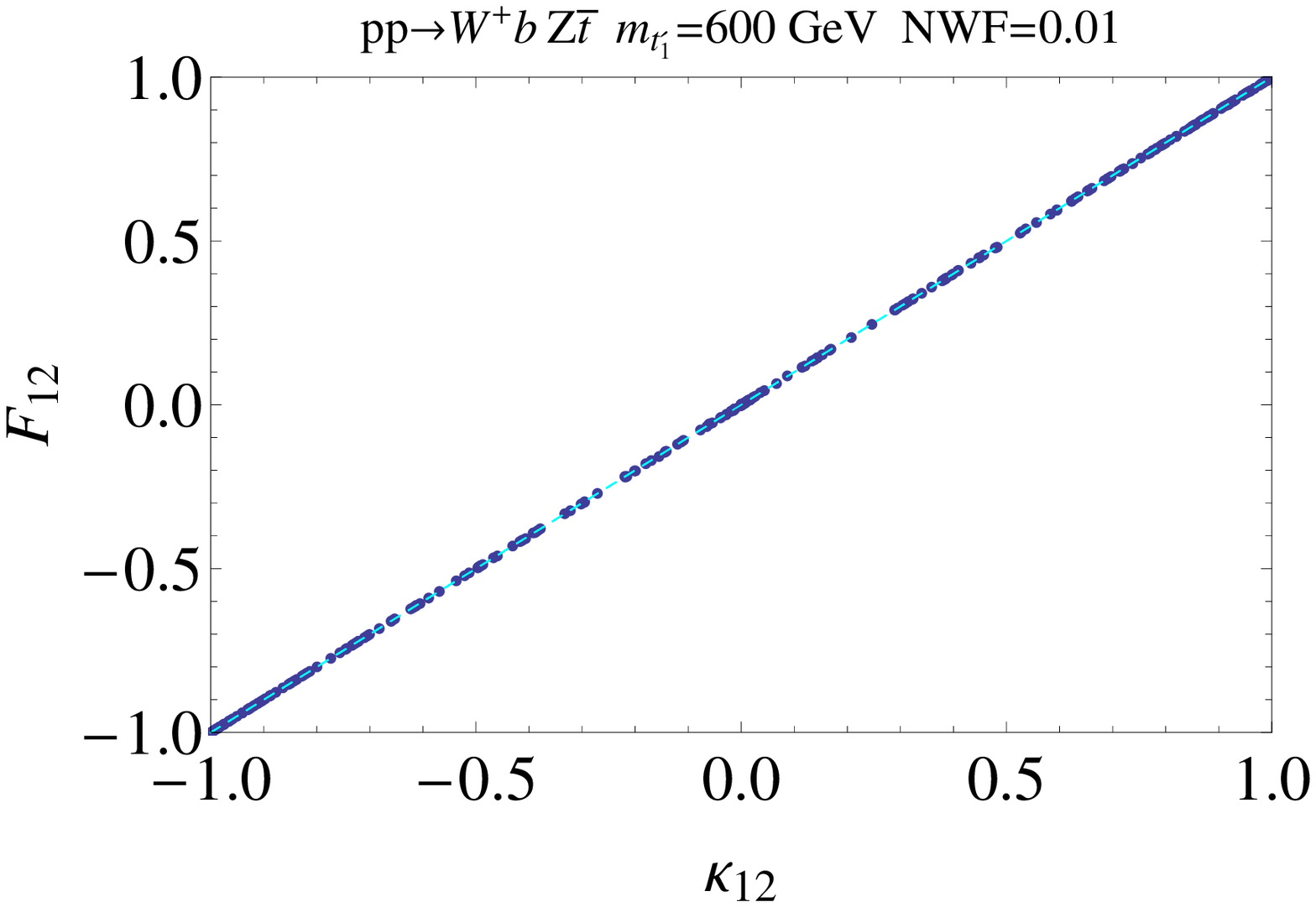, width=0.5\textwidth}%
\epsfig{file=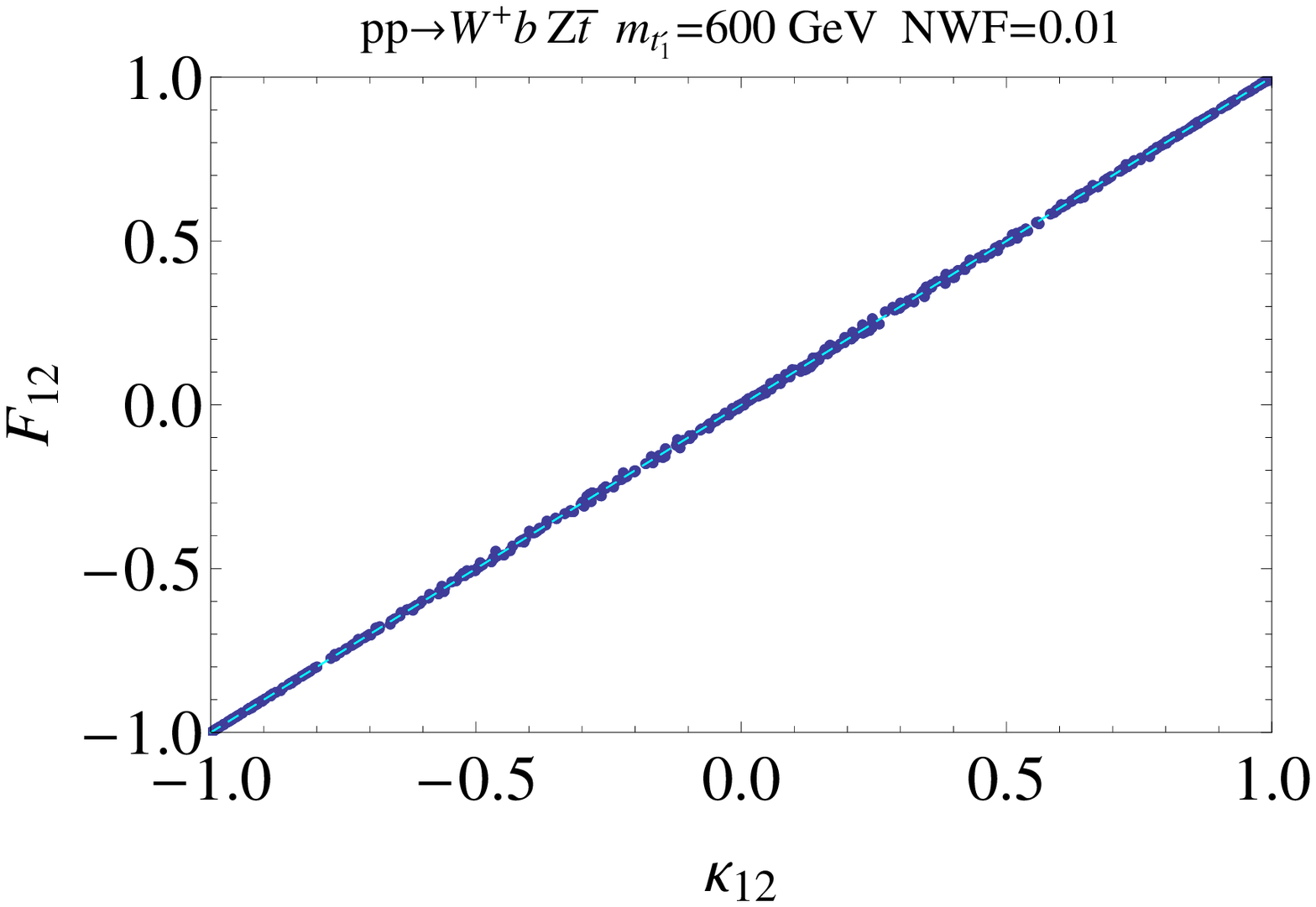, width=0.5\textwidth}\\
\epsfig{file=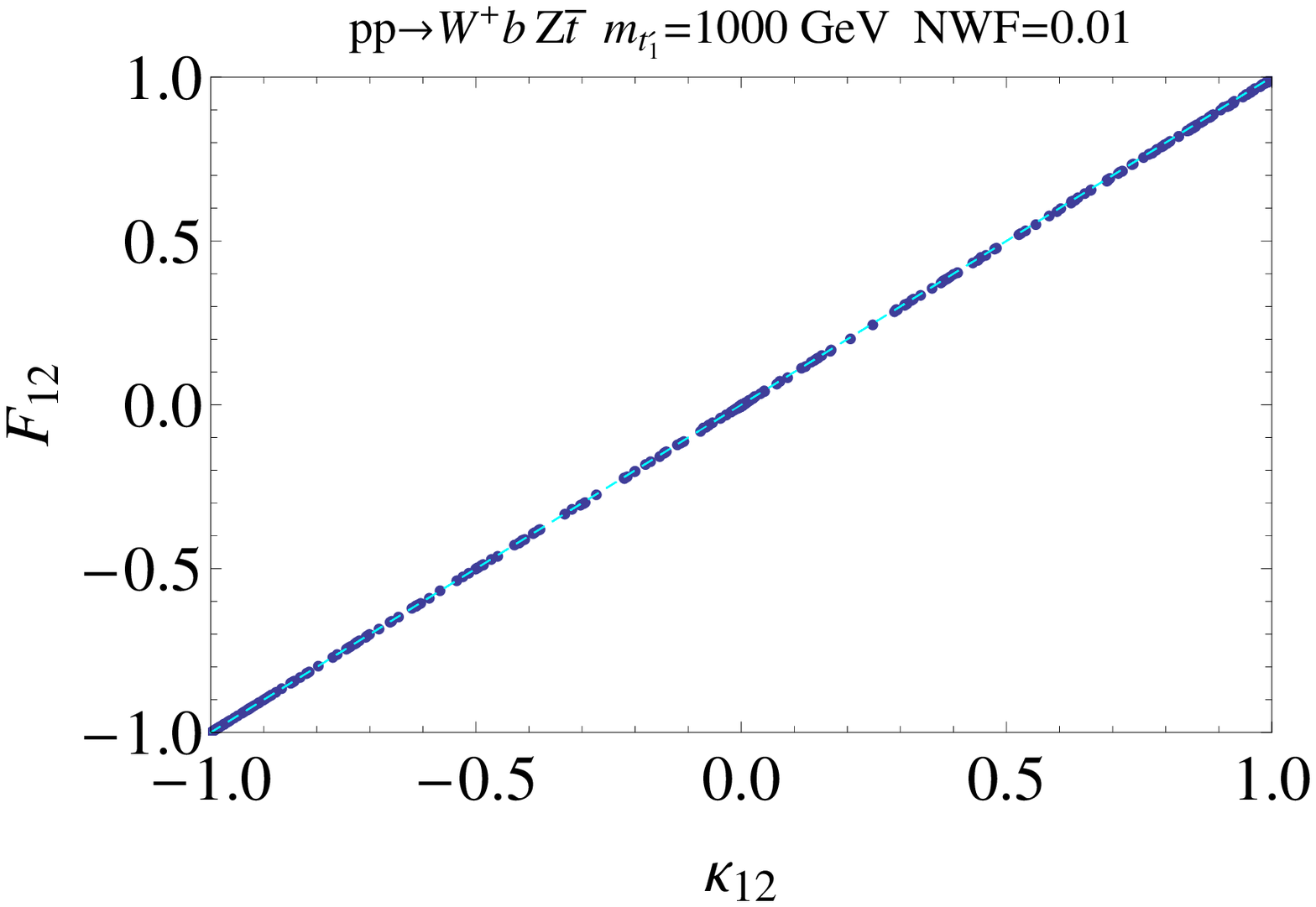, width=0.5\textwidth}%
\epsfig{file=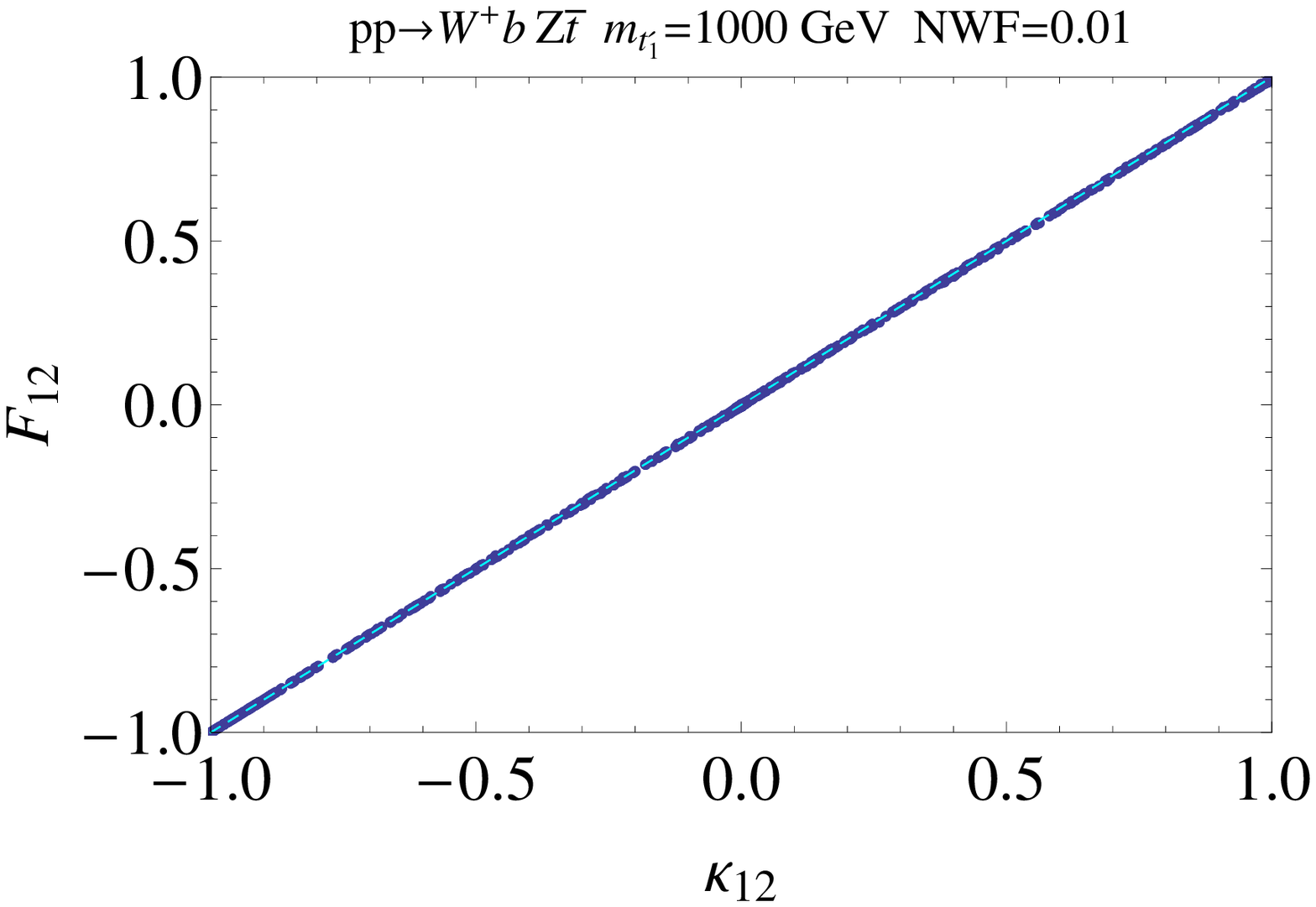, width=0.5\textwidth}
\caption{Interference term $F_{ij}$ as a function of $\kappa_{ij}$. In the left frame the couplings are chiral while in the right one they are general. 
The cyan-dashed line is the bisector in the $\kappa_{ij}-F_{ij}$ plane. Blue points
 are the results of the scan on the couplings for $m_{t_1^\prime}=300,600,1000~$GeV, with different values of the mass splitting between $t_1$ and $t_2$. The Narrow Width Factor (NWF) is the upper limit on max$(\Gamma_{t^\prime_1}$/$m_{t_1^\prime},$ $\Gamma_{t^\prime_2}/$ $m_{t_2^\prime})$ for each point of the scan.}
\label{fig:wbzt_int}
\end{figure}

We have performed a scan on the $Q_V$s couplings for different values of masses and splitting between the two $t^\prime$s and we have obtained the value of the interference term (\ref{orderparameter}) through numerical simulation with MadGraph5~\cite{Alwall:2011uj}
and alternatively cross-checked via CalcHEP3.4~\cite{Belyaev:2012qa}.
 The results are shown in Fig.\ref{fig:wbzt_int} (left frame), where it is possible to notice a remarkable linear correlation between $F_{ij}$ and the expression in Eq.(\ref{eq:int2}).

If the chirality of the couplings of $t^\prime_1$ and $t^\prime_2$ with respect to the SM quarks is opposite, interference effects can arise when the masses of the quarks in the final state are not negligible, as is in the case of decay to top quarks. Considering a scenario where  $t^\prime_1$s decays predominantly to $Zt_L$ and $t^\prime_2$ does so in $Zt_R$, then the interference between $t_L$ and $t_R$ may in principle become relevant. We have numerically verified, however, that in case the chirality of the two $Q_V$ is opposite, the interference effect between massive final states is always negligible, unless the $Q_V$s masses approach the threshold of the final state. This case implies, however, very light $Q_V$s, with masses of the order of 300 GeV, and this range is already excluded by experimental searches.

We show in Fig.\ref{fig:wbzt_int} (right frame) the results for the analogous process (\ref{eq:proc}) where both chiralities are now present in the couplings of $Q_V$s: this process is described by the generalised Eq.(\ref{eq:gen2}). Interference effects between final state quarks of different
chiralities become relevant when the masses of the heavy quarks are close to the top mass, but, as already stressed, this scenario has been tested only to show the appearance of  chirality flipping interference effects, since such a low value for the mass of the heavy quarks is already experimentally excluded.

\subsection{Differential distributions}
The results of the previous sections only apply to the total cross section of the process of pair production and decay of the heavy quarks. However, it is necessary to evaluate how kinematic distributions are affected by the presence of interference terms, as experimental efficiencies of a given search may be largely different if the kinematics of the final state is not similar to the case without
 interference. To evaluate the contribution of interference we have considered the process $p p \to W^+ b Z \bar t$, with subsequent semileptonic decay of the top, mediated by two heavy top-like partners $t_1^\prime$ and $t_2^\prime$ in three limiting cases:
\begin{itemize}
 \item degenerate masses ($m_{t_{1,2}^\prime}=600$ GeV) and couplings with same chirality (both left-handed);
 \item degenerate masses ($m_{t_{1,2}^\prime}=600$ GeV) and couplings with opposite chirality;
 \item non-degenerate masses ($m_{t_1^\prime}=600$ GeV, $m_{t_2^\prime}=1.1m_{t_1^\prime}=660$ GeV) and couplings with same chirality (both 
left-handed).
\end{itemize}
\begin{figure}[htb]
\begin{minipage}{.48\textwidth}
\textbf{Scalar sum of transverse momentum}
\end{minipage}\hfill
\begin{minipage}{.48\textwidth}
\textbf{Missing transverse energy}
\end{minipage}
\vskip 10pt
\epsfig{file=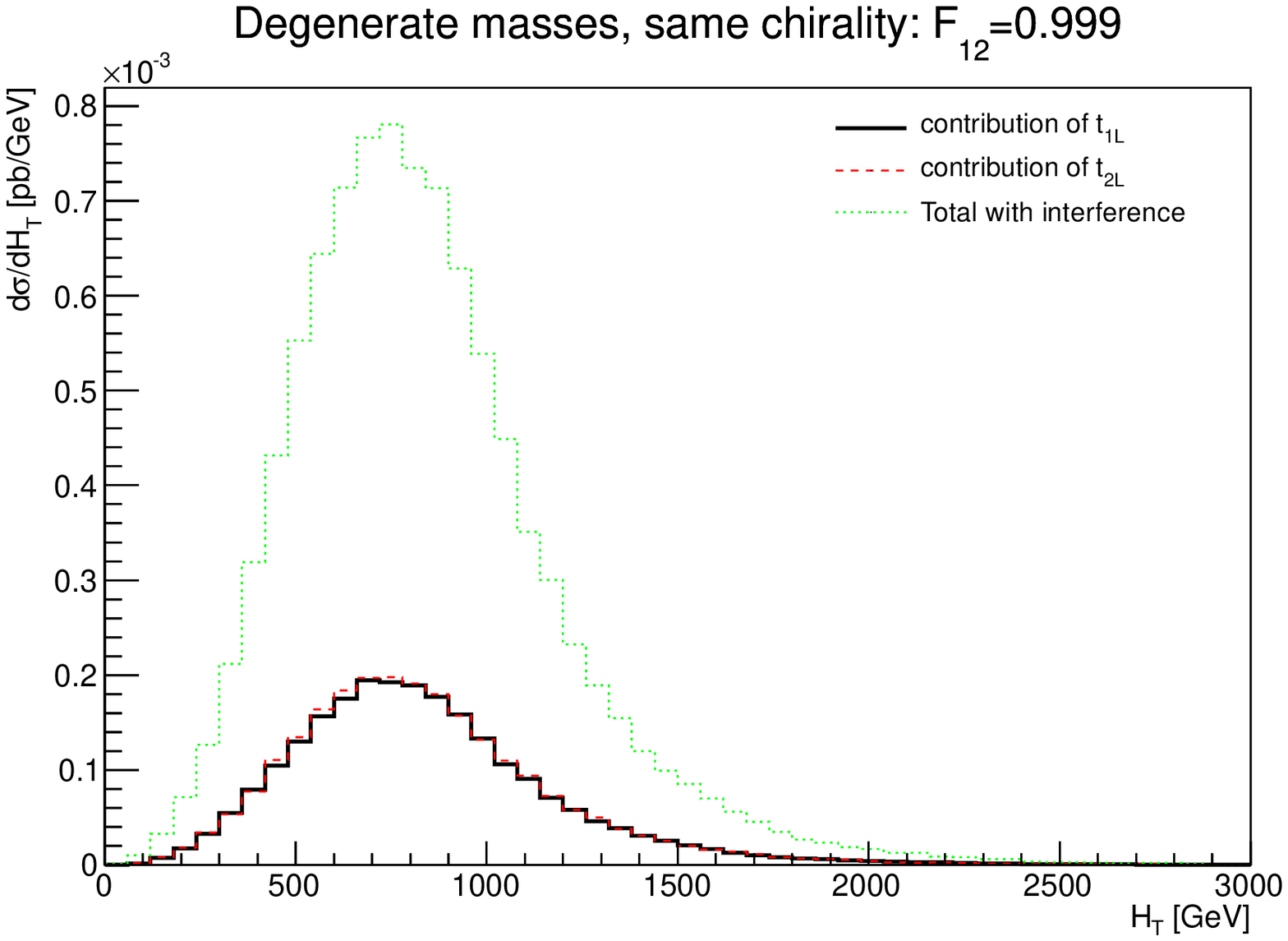, width=0.5\textwidth}%
\epsfig{file=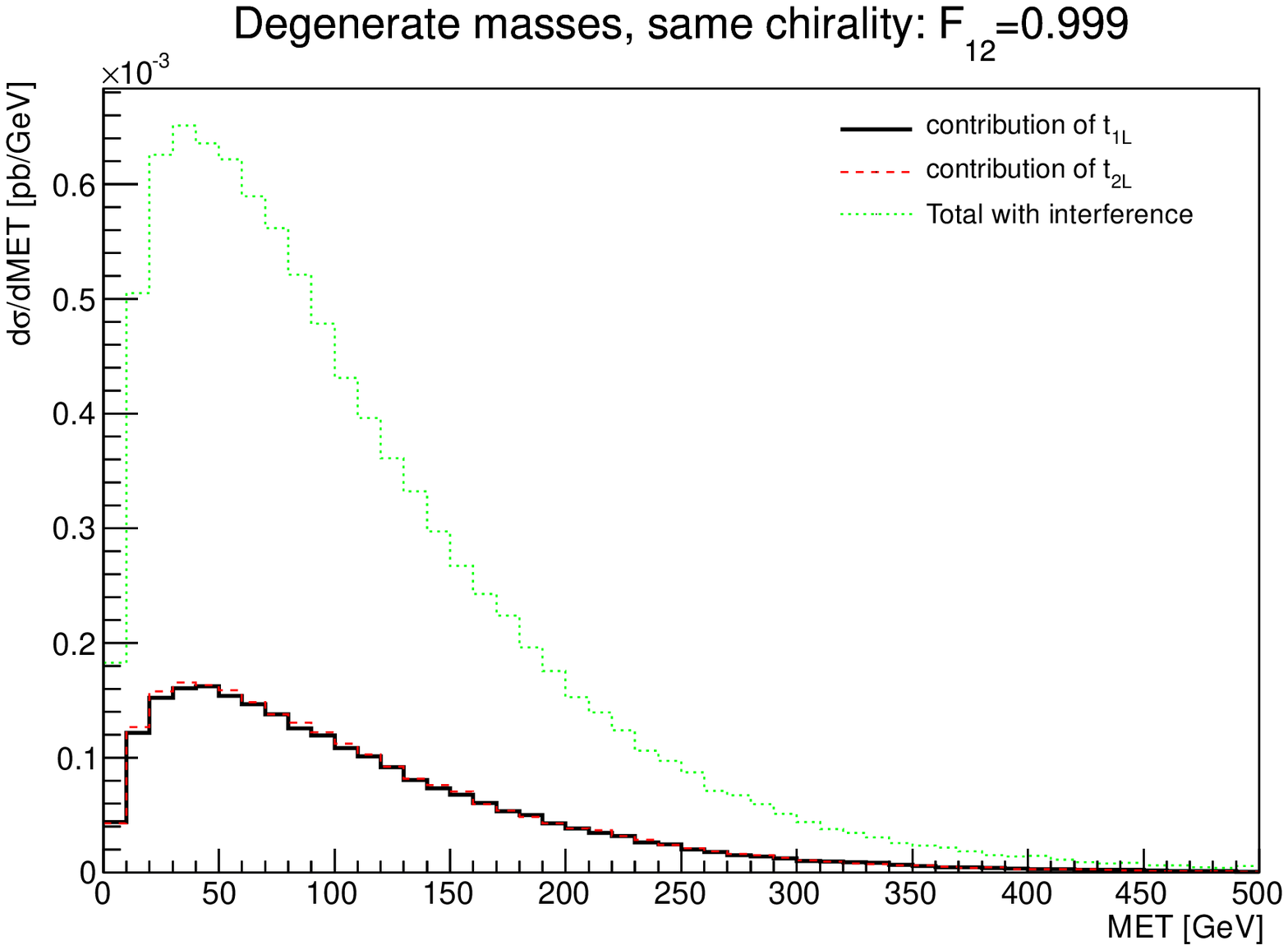, width=0.5\textwidth}\\
\epsfig{file=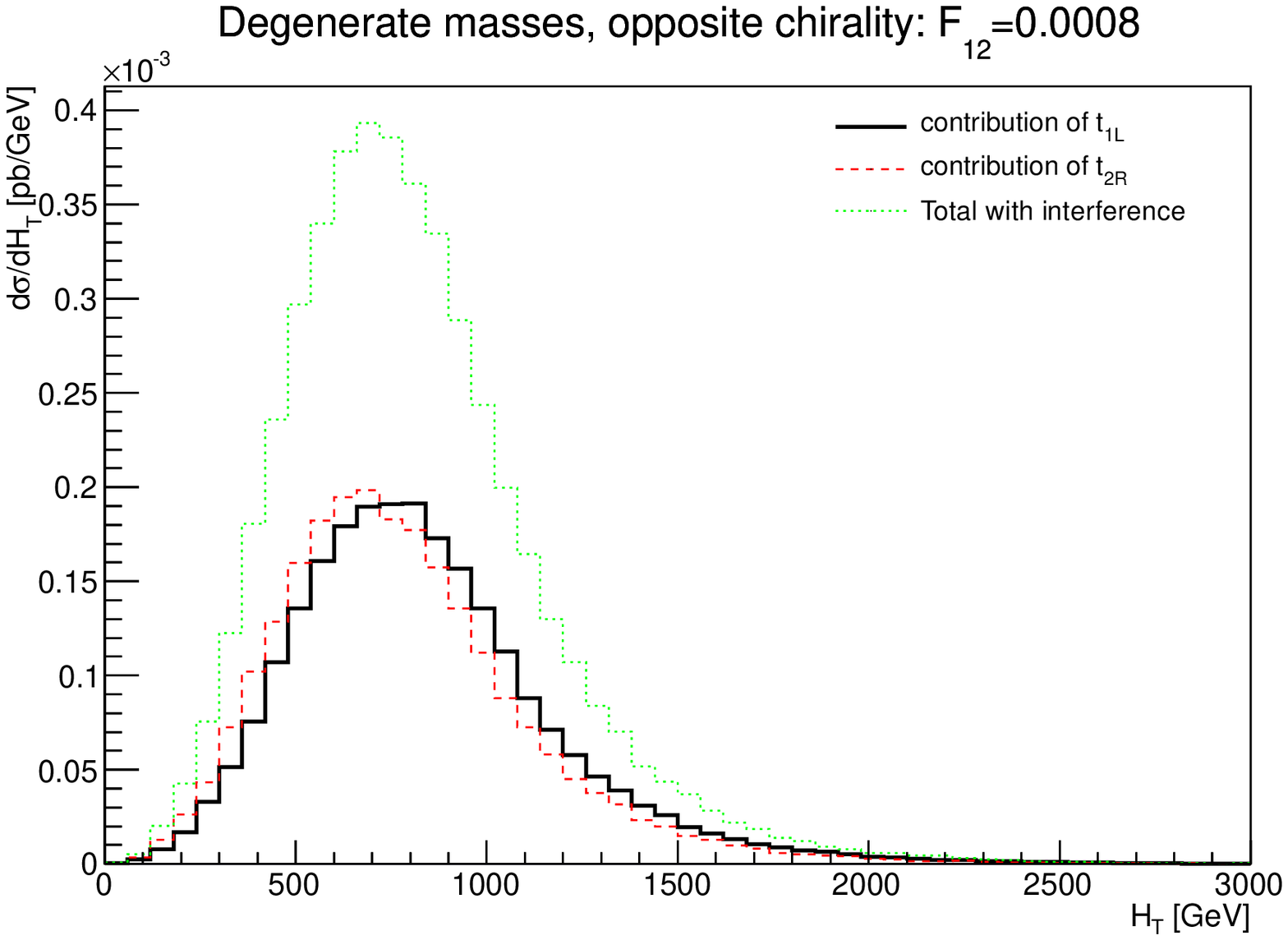, width=0.5\textwidth}%
\epsfig{file=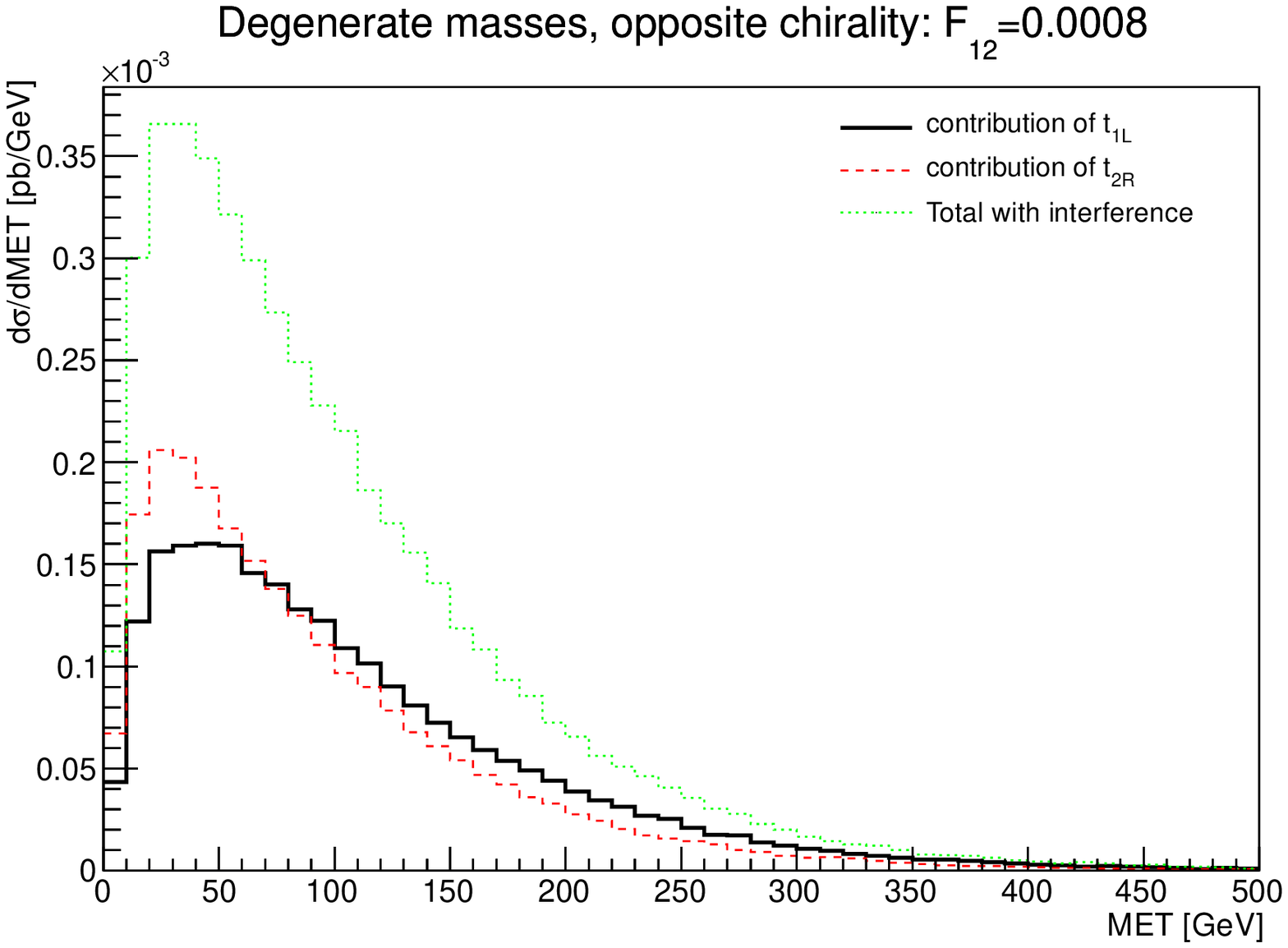, width=0.5\textwidth}\\
\epsfig{file=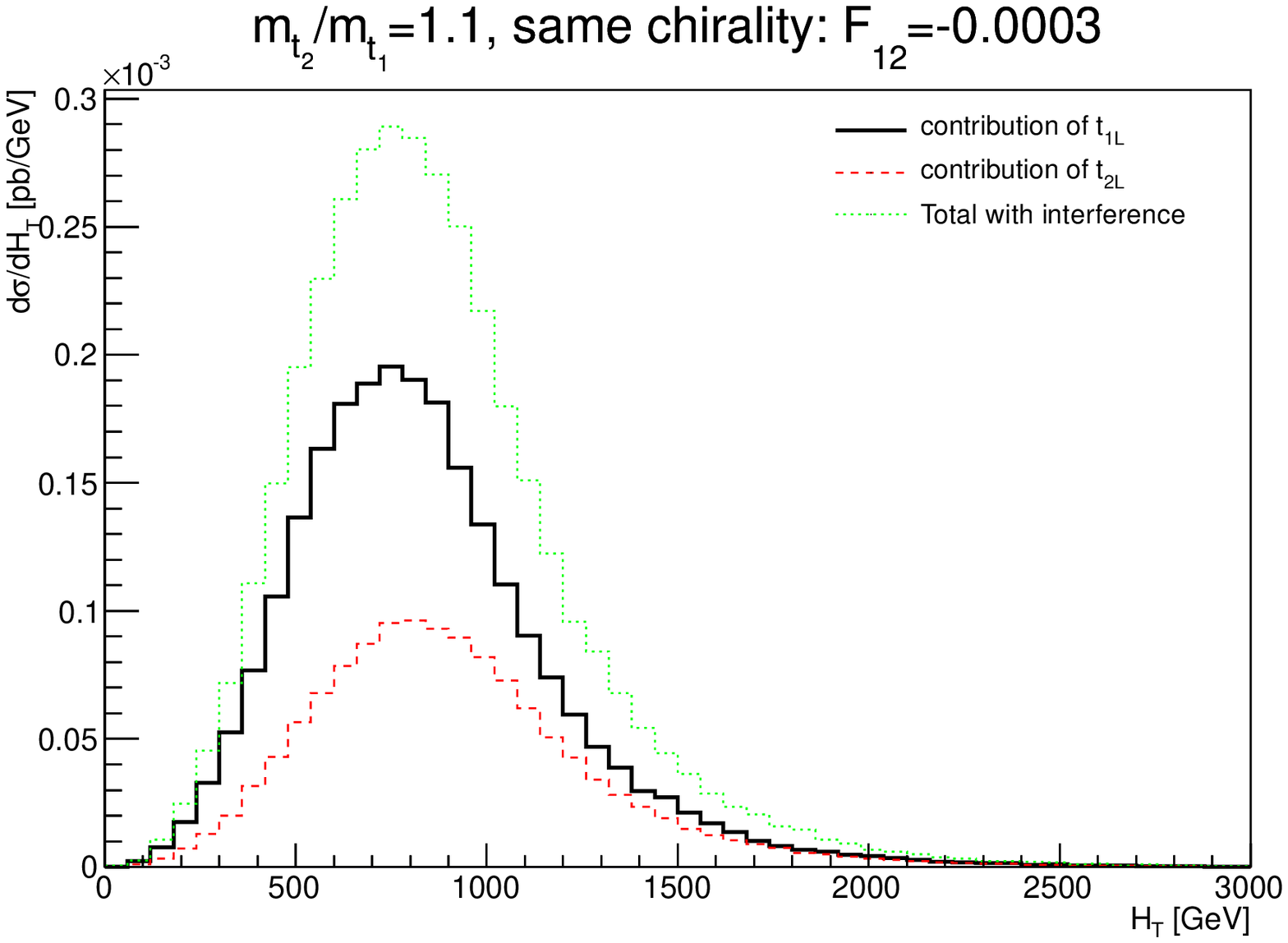, width=0.5\textwidth}%
\epsfig{file=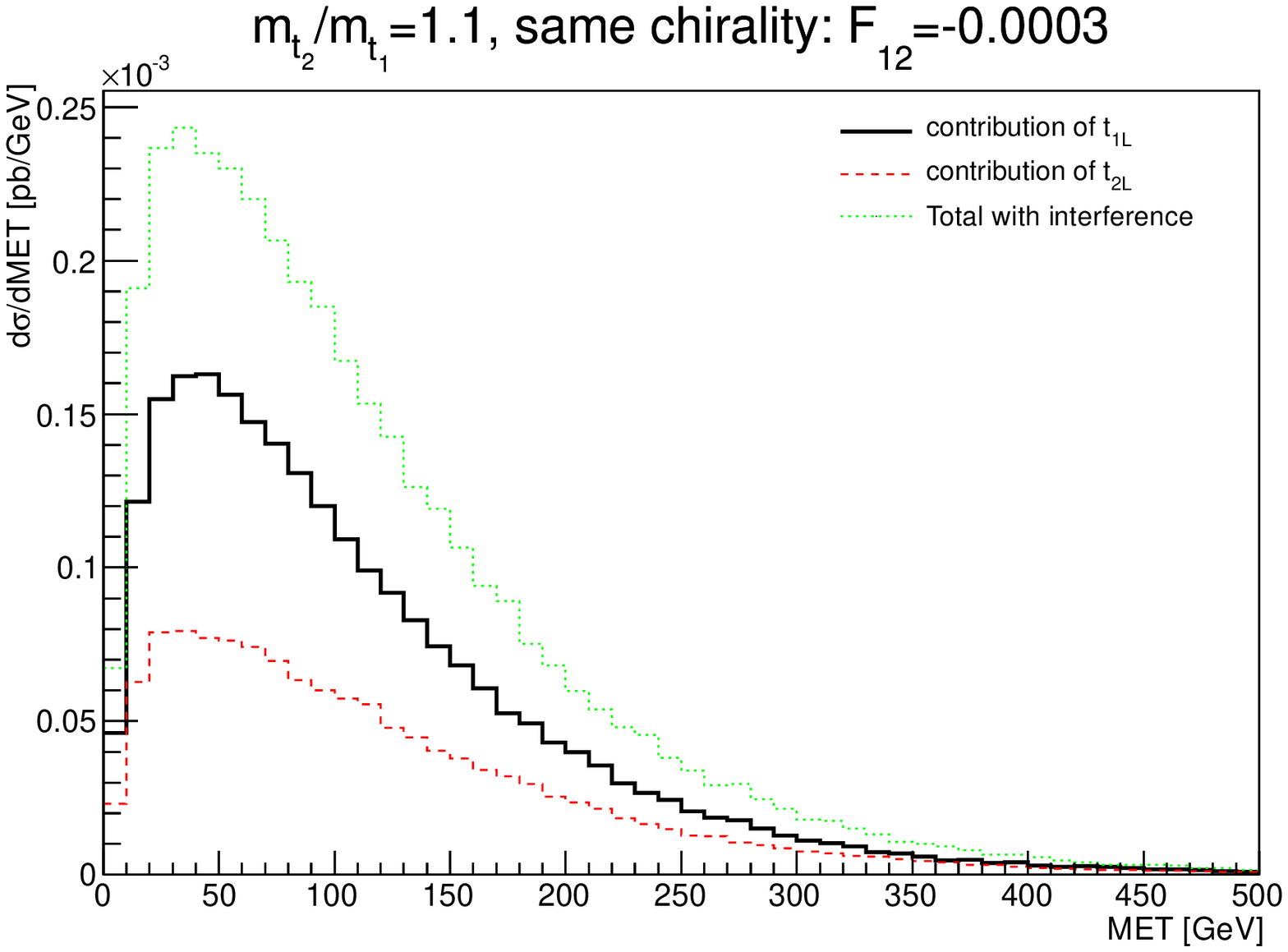, width=0.5\textwidth}
\caption{Differential distributions for $H_T$ and $\MET$ for the process $pp\to W^+bZ\bar t\to W^+bZ \bar b e^- \bar \nu_e$ in three different scenarios: degenerate masses and couplings with same chirality (top); degenerate masses and couplings with opposite chirality (middle); non-degenerate masses ($m_{t_2^\prime}=1.1m_{t_1^\prime}$) and couplings with same chirality (bottom). Here, $m_{t_1^\prime}$ has been fixed to 600~GeV. The values of the interference term $F_{12}$ are shown for each scenario.}
\label{fig:wbzt_dist}
\end{figure}

The results are shown in Fig.\ref{fig:wbzt_dist}, where we display the $H_T$ (scalar sum of the transverse momenta of jets) and $\MET$ (missing transverse energy) differential distributions. When the interference is maximal, all distributions have exactly the same features, that  is, the distributions including interference can be obtained by a rescaling of the distributions for production of the two heavy quarks using $(1+\kappa_{ij})$ for the rescaling factor: this relation comes from considering Eq.~(\ref{orderparameter}) and the linear correlation between $F_{ij}$ and $\kappa_{ij}$ verified in the previous section. Therefore our results for the total cross section can also be applied at differential level and, specifically, it is possible to apply
 the same experimental efficiencies to the case of a single heavy quark or to the case with degenerate quarks with couplings of identical chirality. 
In contrast, in the two other scenarios we have considered, where interference is negligible,
 the distributions for production of either $t_1^\prime$ or $t_2^\prime$ exhibit different features and the distribution of the total process is, for each bin, simply the sum of the distributions of the two heavy quarks (i.e. the rescaling factor is 1 because $k_{ij}\sim0$). Same patterns are seen for all other differential distributions that we have investigated: (pseudo)rapidity, cone separation, etc.

\begin{figure}[htb]
\epsfig{file=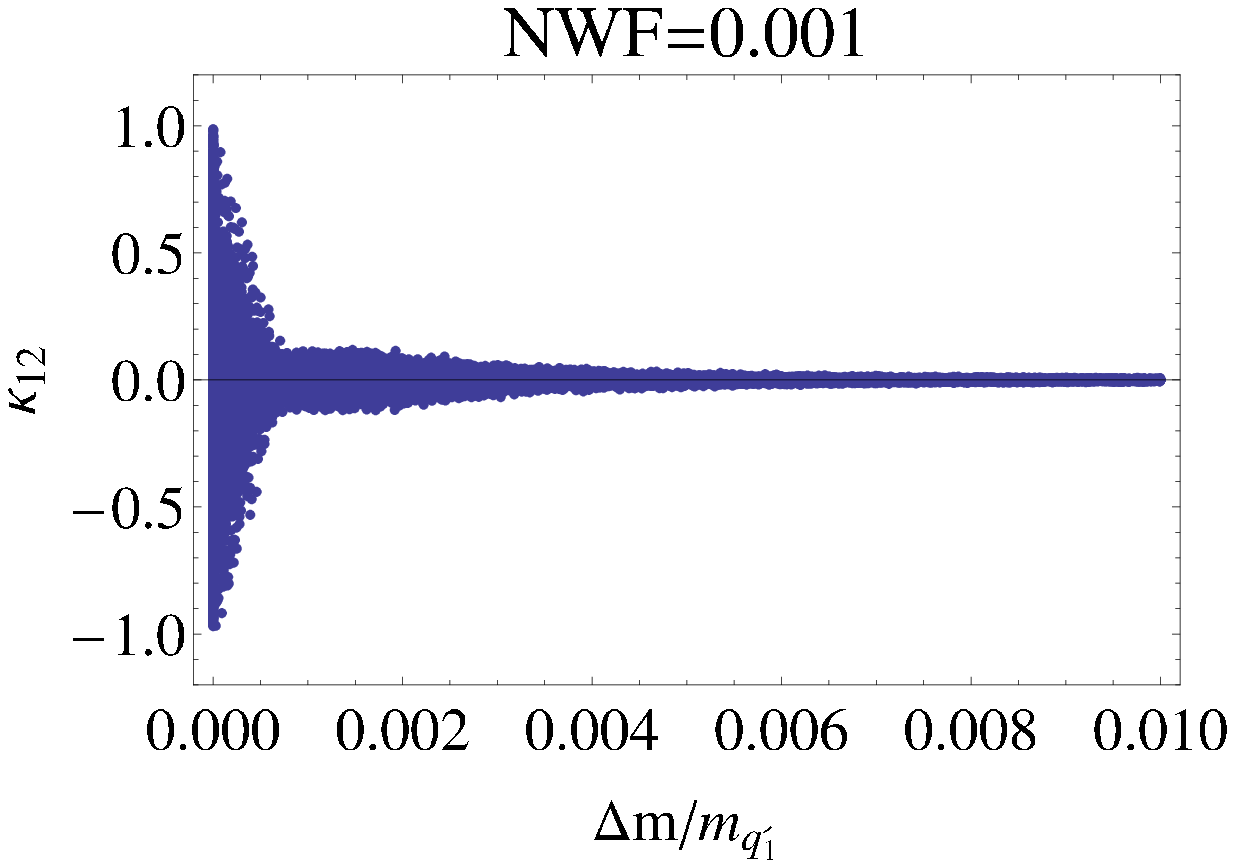, width=.48\textwidth}\hfill
\epsfig{file=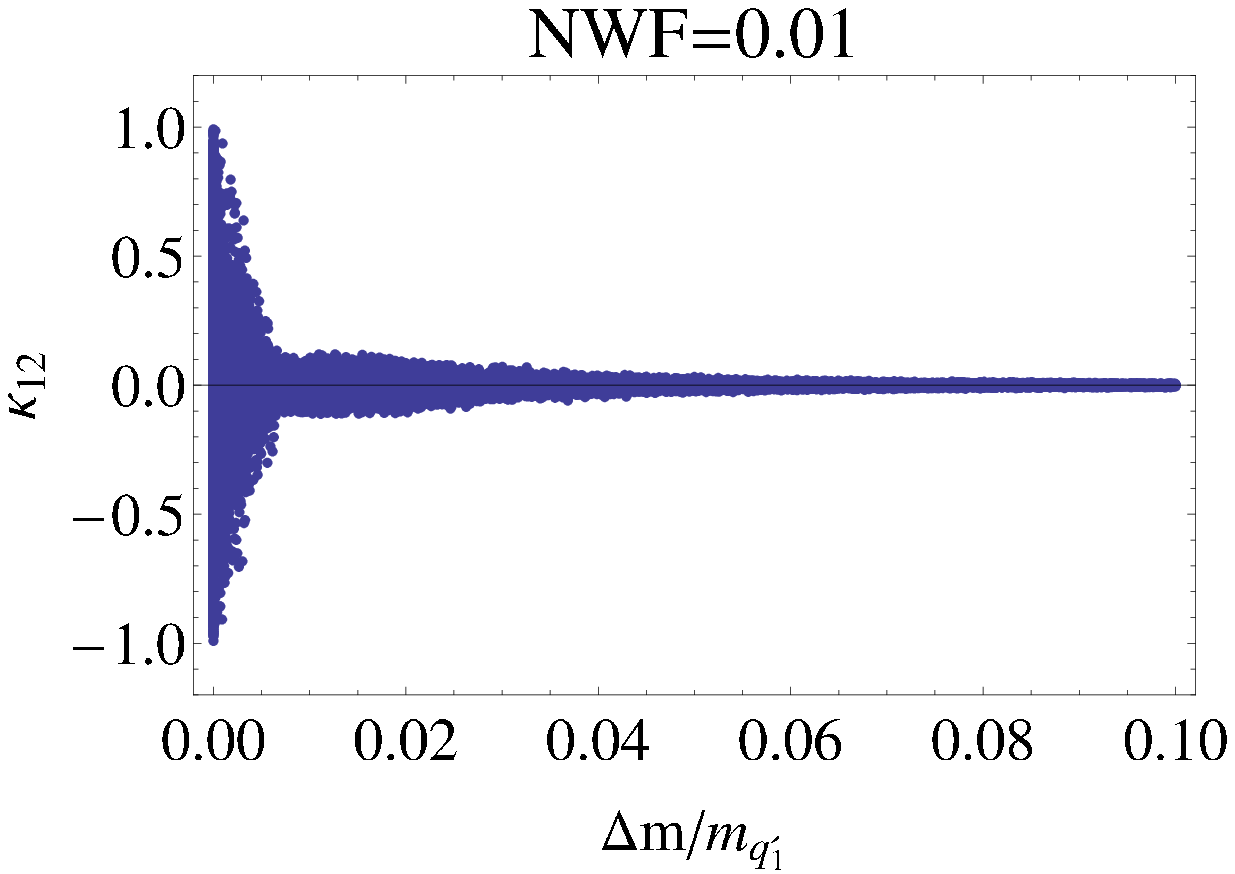, width=.48\textwidth}
\caption{The range of the interference contributions with respect to the mass splitting between the heavy quarks for different values of the NWF. 
Notice the different scales of the $x$ axis.}
\label{fig:intsplit}
\end{figure}

As a final remark, we may ask how much the range of the possible values for the interference term drops by increasing the mass splitting between the heavy quarks and, therefore, when should we consider the interference as always negligible. In Fig.\ref{fig:intsplit} it is possible to notice that the range of values for the parameter $\kappa_{12}$ drops extremely fast with the mass splitting and depends on the value of the NWF. The range of the interference contributions, however, becomes smaller than 10\% in a region of mass splitting where the shapes of the distributions can be safely considered as equivalent.

\subsection{Validity range of the model-independent approach and ``master formula" for the interference}

In this subsection we  discuss the range of validity of the analytical formula for $\kappa_{ij}$ describing the interference
effect. Our ansatz was made under the assumption of small $\Gamma/m$ ratios, which, in terms of probability (e.g. amplitude square), means that the QCD production part of the $Q_V$s and their subsequent decay can be factorised.
We then took advantage of this consideration by making this factorisation already at amplitude level and writing therefore the interference, Eq.(\ref{eq:intpart}), and pair production, Eq.(\ref{eq:prodpart}), contribution to the total cross section as a modulus squared of quantities that do not involve the QCD production part, then using then these two relations to define our $\kappa_{ij}$ parameter in Eq.(\ref{eq:int1}).
This concept of factorisation is valid just in the limit $\Gamma/m \to 0$, for which, however, there will be no decay of the $Q_Vs$ and therefore no interference at all. It is nonetheless clear that this approximation of factorisation of production and decay will be the more accurate the more this ratio is closer to zero. In fact, in the previous subsections we have shown that the formula for $\kappa_{ij}$ reproduces the true interference $F_{ij}$ very accurately in the case of NWF=$\Gamma/m=0.01$. It is however very informative to explore the range of validity of our ansatz in function of the NWF parameter, especially in view of practical applications of our method.

\begin{figure}[htb]
\epsfig{file=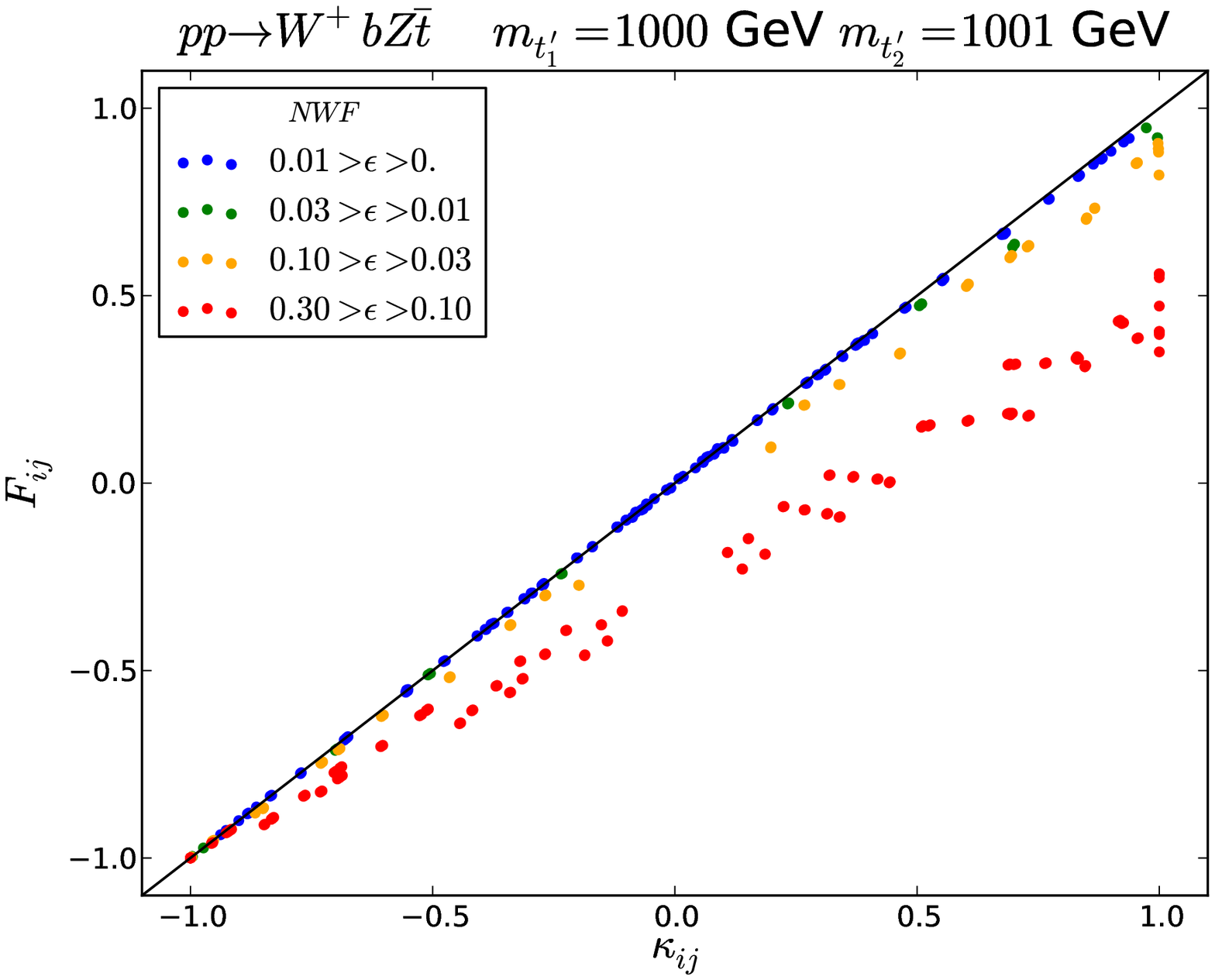,width=0.5\textwidth}%
\epsfig{file=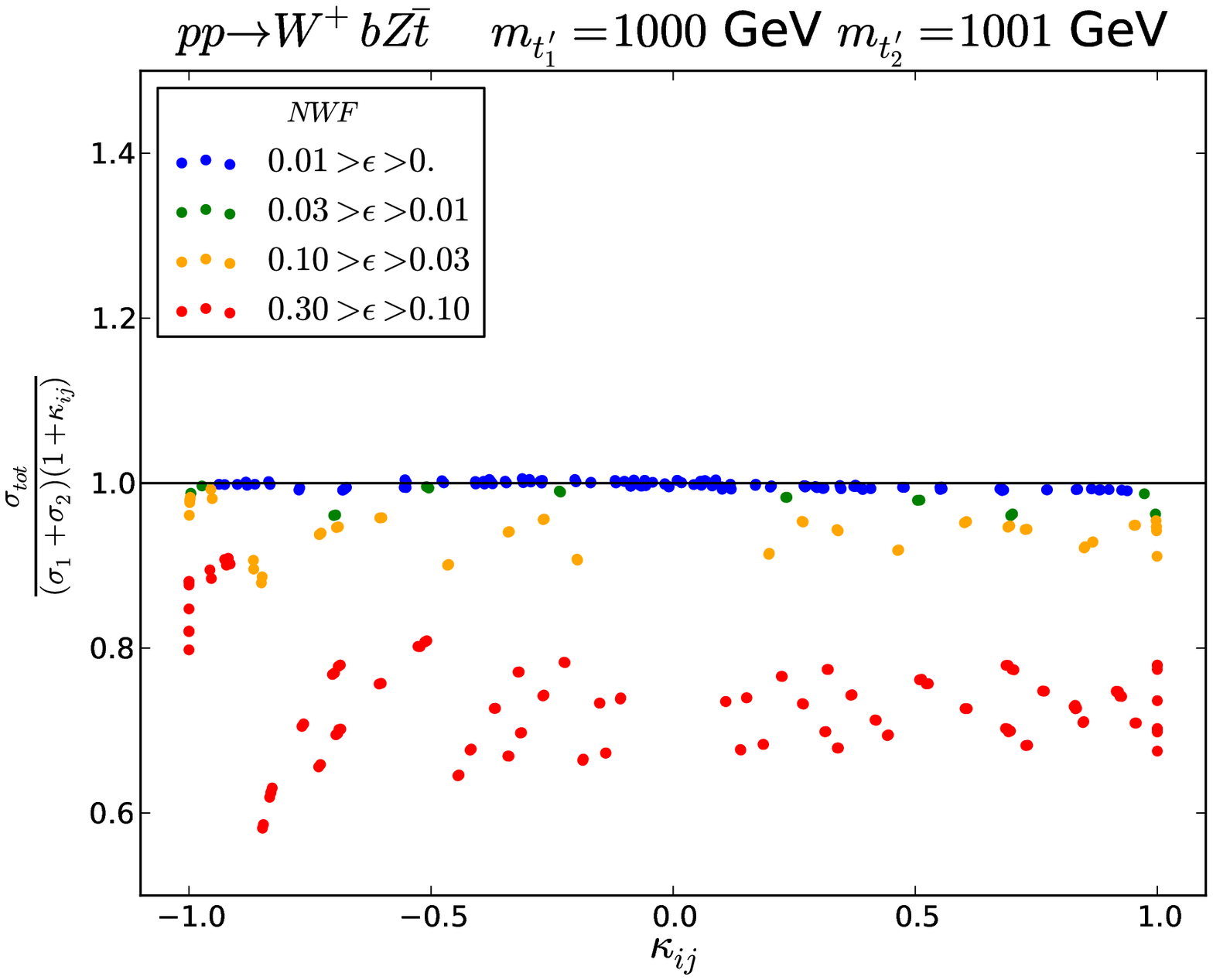,width=0.5\textwidth}%
\caption{$F_{ij}$ versus $\kappa_{ij}$ (left)
and  $\frac{\sigma_{tot}}{(\sigma_1+\sigma_2)(1+\kappa_{ij})}$  versus $\kappa_{ij}$ (right)
for various values of  
the NWF for the $pp\to W^+bZ\bar{t}$ process.
\label{fig:range-of-validity}}
\end{figure}

In Fig.~\ref{fig:range-of-validity} (left)
we present results for $F_{ij}$ versus $\kappa_{ij}$ 
for values of the NWF in the  0.0--0.3  range for the $pp\to W^+bZ\bar{t}$
process. One can see that our description of the interference remains at a
quite accurate level for NWF below about $10\%$ while already in the range 10\%--30\%
one can see non-negligible deviations from the analytic formula predictions, i.e., $\kappa_{ij}$,
as compared to the true value of the interference, $F_{ij}$.
The ``triangle" shape of the pattern of the left frame of Fig.~\ref{fig:range-of-validity}
is simply related to the fact that,
in case of large negative interference, the
${\sigma^{\rm tot}_{ij}}$ value is close to zero. Therefore, even in case of 
large $relative$ deviations, the $predicted$ value of   ${\sigma^{\rm tot}_{ij}}$
will be still close to zero, forcing $F_{ij}$ to be around $-1$, according to 
Eq.~(\ref{orderparameter}), even in case of large values of the NWF  parameter.
Therefore, it is important to look at the complementary plot presenting
$\frac{\sigma_{tot}}{(\sigma_1+\sigma_2)(1+\kappa_{ij})}$  versus $\kappa_{ij}$
shown in Fig.~\ref{fig:range-of-validity} (right).
One can see that deviations of the  cross-section predicted by the ``master formula",
$(\sigma_1+\sigma_2)(1+\kappa_{ij})$,
from the real one, $\sigma_{tot}$, depends only on the value of NWF. For large values of NWF
one can also see 
that $\sigma_{\rm tot}$ is below $(\sigma_1+\sigma_2)(1+\kappa_{ij})$,
which is related to the fact that in case of  $\sigma_{\rm tot}$ the pure Breit-Wigner shape of the $t'_i$
resonances is actually distorted and suppressed on the upper end due  to steeply falling parton distribution functions.
Furthermore, one should note that the quite accurate description of the interference found at the integrated level for
 NWF $<0.1$ remains true at differential level too.
Finally, we remark that the multi-parametric scan was done using CalcHEP3.4 on the HEPMDB database~\cite{hepmdb}, where the model studied here can be found 
 under the 
\url{http://hepmdb.soton.ac.uk/hepmdb:1113.0149} link.

\section{Conclusions}

We have studied the role of interference in the process of pair production of new heavy (vector-like) quarks.
Considering such interference effects is crucial for the reinterpretation of the results of experimental searches of new quarks decaying to the same final state in the context of models with a new quark sector, which is usually not limited to the presence of only one heavy quark.
We have shown that, if the small $\Gamma/m$ approximation holds, and therefore it is possible to factorise the production and decay of the new quarks, the interference contribution can be described by considering a parameter which contains only the relevant couplings and the scalar part of the propagators of the new quarks. 

We have obtained a remarkably accurate description of the
exact interference (described by the  term $F_{12}$ defined in Eq.~(\ref{orderparameter}))
using a simple analytical formula for the parameter $\kappa_{ij}$ defined in Eq.(\ref{eq:int2}).
This description  holds regardless of the chiralities of the couplings between the new and SM quarks, Eq.(\ref{eq:gen2}).
This means that it is possible to analytically estimate, with very good accuracy, the interference contribution to the pair production of two (and possibly more) quarks pairs decaying into the same final state, once couplings, total widths and masses are known, without performing a dedicated simulation or a full analytical computation.
We have also discussed the region of validity of this approximation in connection to 
the mixing effects at the loop-level contribution to a heavy quark self-energy
which could potentially lead to a non-negligible interference. Therefore,
in order to use the analytical formula for the interference we have derived,
one should verify that  the off-diagonal contributions to the propagators  are suppressed
and  check that the relation analogous to Eq.(18) takes place for the particular model
under study.

We have verified that also at the level of differential distributions it is possible to obtain the distributions including interference by a simple rescaling of those of the heavy quarks decaying to the given final state. Finally, we have checked that the linear correlation does not hold anymore for large values of the
 $\Gamma/m$ ratio, while it has been verified that for a NWF less than 10\% (which is very typical for all classes of models with $Q_V$s), 
the expressions for $\kappa_{ij}$ do indeed provide an accurate description of the interference term.
When interference effects are relevant and in the range of validity of our expressions, it is therefore possible to apply the same experimental efficiencies used for individual quark pairs to the full process of production and decay of two pairs of new quarks.

\section*{Acknowledgments}
The authors would like to thank M. Buchkremer, G. Cacciapaglia, A. Deandrea and S. De Curtis for useful discussions. 
They also thank A. Pukhov for a quick response which helped to improve  the  interference treatment within CalcHEP.
DB, AB, SM and LP are financed in part through the NExT Institute. DB and LP would like to thank the Galileo Galilei Institute (GGI) in Florence for hospitality while part of this work was carried out. JB thanks the University of Southampton  for the Summer Student programme support.
AB and JB acknowledge the use of the HEPMDB and IRIDIS HPC Facility at the University of Southampton in the completion of this study.

\end{document}